# Diversifying Top-K Results

Lu Qin, Jeffrey Xu Yu, Lijun Chang
*The Chinese University of Hong Kong, Hong Kong, China*
{lqin,yu,ljchang}@se.cuhk.edu.hk

## ABSTRACT

Top-$k$ query processing finds a list of $k$ results that have largest scores w.r.t the user given query, with the assumption that all the $k$ results are independent to each other. In practice, some of the top-$k$ results returned can be very similar to each other. As a result some of the top-$k$ results returned are redundant. In the literature, diversified top-$k$ search has been studied to return $k$ results that take both score and diversity into consideration. Most existing solutions on diversified top-$k$ search assume that scores of all the search results are given, and some works solve the diversity problem on a specific problem and can hardly be extended to general cases. In this paper, we study the diversified top-$k$ search problem. We define a general diversified top-$k$ search problem that only considers the similarity of the search results themselves. We propose a framework, such that most existing solutions for top-$k$ query processing can be extended easily to handle diversified top-$k$ search, by simply applying three new functions, a sufficient stop condition sufficient(), a necessary stop condition necessary(), and an algorithm for diversified top-$k$ search on the current set of generated results, div-search-current(). We propose three new algorithms, namely, div-astar, div-dp, and div-cut to solve the div-search-current() problem. div-astar is an A* based algorithm, div-dp is an algorithm that decomposes the results into components which are searched using div-astar independently and combined using dynamic programming. div-cut further decomposes the current set of generated results using cut points and combines the results using sophisticated operations. We conducted extensive performance studies using two real datasets, enwiki and reuters. Our div-cut algorithm finds the optimal solution for diversified top-$k$ search problem in seconds even for $k$ as large as $2,000$.

## 1. INTRODUCTION

Top-$k$ queries are one of the most fundamental queries used in the IR and database areas. Given a user query, the top-$k$ results of the query are a list of $k$ results that have largest scores/relevances with respect to the user query, under the assumption that all of the $k$ results are independent to each other. In some situations, for a certain top-$k$ query, some of the results returned can be very similar to each other. For example, if we search "apple" in Google image[1], 7 out of the top-10 results returned are the logo of the Apple company. In order to remove the redundancy in the results, and at the same time keep the quality of the top-$k$ results, diversity should be considered in the top-$k$ search problems.

For top-$k$ search algorithms. In the literature, most of them aim at finding an early stop condition, such that they can find the top-$k$ results without exploring all the possible search results. Based on this, two frameworks are generally used, namely, the incremental top-$k$ framework and the bounding top-$k$ framework. The incremental top-$k$ framework outputs the results one by one in non-increasing order of their scores, and stops as soon as $k$ results are generated. It aims to find a polynomial delay algorithm such that given the existing generated results, the next result with largest score can be generated in polynomial time w.r.t. the size of the input only [16, 15, 20, 14]. In the bounding top-$k$ framework, results are not necessarily generated in non-increasing order of their scores. It maintains a score upper bound for the unseen results every time when a new result is generated. The algorithm stops when the current $k$-th largest score is no smaller than the upper bound for the unseen results. The threshold algorithm based approaches [7, 9] fall in this framework and other approaches include [12, 17].

Diversity aware search has been studied in recent years. Most of the existing solutions that support diversity on top-$k$ search results assume the ranking of all the search results are given in advance. Based on which, a diversity search algorithm is given to output $k$ results based on a scoring function that takes both query relevance and diversity into consideration [6, 1, 11, 5, 2]. Other works give algorithms that solve the diversity problem for a special area, i.e., graph search [18], document search [22], etc. and can hardly be extended to support general top-$k$ diversity search.

In this paper, we propose a general framework to handle the diversified top-$k$ search problem. We keep the advantages for the existing top-$k$ search algorithms, that can stop early without exploring all search results, and at the same time, we take diversity into consideration. We show that any top-$k$ search algorithm that can be used in the incremental top-$k$ framework or the bounding top-$k$ framework can be easily extended to handle diversified top-$k$ search, by adding three new functions studied in this paper: a sufficient stop condition sufficient(), a necessary stop condition necessary(), and a diversity search function div-search-current(). All of them are application independent. The only assumption in our framework is that, given any two search results $v_i$ and $v_j$, whether $v_i$ and $v_j$ are similar to each other can be decided, e.g., using a similarity function $\mathsf{sim}(v_i, v_j) > \tau$ for a user given threshold $\tau$. We output a list of $k$ results with maximum total scores such that no two of them are similar to each other. We make the following contributions in this paper.

[1] http://www.google.com/imghp

(1) We formalize the diversified top-$k$ search problem. Based on our definition, the optimal solution only depend on the similarity of search results themselves, and no other information is needed.

(2) We study two categories of algorithms generally used in finding top-$k$ results with early stop in the literature, namely, the incremental top-$k$ framework and the bounding top-$k$ framework. We show both frameworks can be extended to diversified top-$k$ search by simply adding three application independent functions studied in this paper, namely, a sufficient stop condition $\mathsf{sufficient}()$, a necessary stop condition $\mathsf{necessary}()$, and a diversity search function $\mathsf{div\text{-}search\text{-}current}()$. The sufficient stop condition helps to early stop and the necessary stop condition helps to reduce the number of $\mathsf{div\text{-}search\text{-}current}()$ processes, since $\mathsf{div\text{-}search\text{-}current}()$ is usually a costly operation.

(3) We show that $\mathsf{div\text{-}search\text{-}current}()$ is an NP-Hard problem and is hard to be approximated. We propose three new algorithms, namely, div-astar, div-dp, and div-cut, to find the optimal solution for $\mathsf{div\text{-}search\text{-}current}()$. div-astar is an A$^*$ based algorithm and is slow to handle a large number of results. div-dp decomposes the results into disconnected components in order to reduce the graph size to be searched using div-astar. Results in div-dp are combined using dynamic programming. div-cut further decomposes each component into several subgraphs to form a cptree, based on the cut points of each component. A tree based search is applied on cptree to find the optimal solution.

(4) We conducted extensive performance studies using two real datasets, to test the performance of the three algorithms. Our div-cut approach can find the diversified top-$k$ results within seconds when $k$ is as large as $2,000$.

The rest of this paper is organized as follows. Section 2 formally defines the diversified top-$k$ search problem. Section 3 shows the two existing frameworks on general top-$k$ search problems. Section 4 shows how to extend the two categories of top-$k$ search approaches to solve diversified top-$k$ search, by defining a sufficient stop condition $\mathsf{sufficient}()$, a necessary stop condition $\mathsf{necessary}()$, and a diversified top-$k$ search algorithm $\mathsf{div\text{-}search\text{-}current}()$ to search on the current result set. Section 5, 6, and 7 give three algorithms to solve the $\mathsf{div\text{-}search\text{-}current}()$ problem. We show our experimental results in Section 8, and introduce the related work in Section 9. Finally, we conclude our paper in Section 10.

## 2. PROBLEM DEFINITION

We consider a list of results $S = \{v_1, v_2, \cdots\}$. For each $v_i \in S$, the score of $v_i$ is denoted as $\mathsf{score}(v_i)$. For any two results $v_i \in S$ and $v_j \in S$, there is a user defined similarity function $\mathsf{sim}(v_i, v_j)$ denoting the similarity between the two results $v_i$ and $v_j$. Without loss of generality, we assume $0 \leq \mathsf{sim}(v_i, v_j) \leq 1$ for any two results $v_i \in S$ and $v_j \in S$, and $\mathsf{sim}(v, v) = 1$ for any $v \in S$. Given an integer $k$ where $1 \leq k \leq |S|$, the top-$k$ results of $S$ is a list of $k$ results $S_k$ that satisfy the following two conditions.

1) $S_k \subseteq S$ and $|S_k| = k$.

2) For any $v_i \in S_k$ and $v_j \in S - S_k$, $\mathsf{score}(v_i) \geq \mathsf{score}(v_j)$.

Here, $S - S_k$ is the set of results that are in $S$ but not in $S_k$, i.e., $S - S_k = \{v | v \in S, v \notin S_k\}$.

Given two results $v_i \in S$ and $v_j \in S$, $v_i$ is similar to $v_j$ iff $\mathsf{sim}(v_i, v_j) > \tau$ where $\tau$ is a user defined threshold, and $0 < \tau \leq 1$. We use $v_i \approx v_j$ to define that $v_i$ is similar to $v_j$.

**Definition 1 (Diversified Top-$k$ Results)** *Given a list of search results $S = \{v_1, v_2, \cdots\}$, and an integer $k$ where $1 \leq k \leq |S|$, the diversified top-$k$ results of $S$, denoted as $D(S)$, is a list of results that satisfy the following three conditions.*

*1) $D(S) \subseteq R$ and $|D(S)| \leq k$.*

*2) For any two results $v_i \in R$ and $v_j \in R$ and $v_i \neq v_j$, if $v_i \approx v_j$, then $\{v_i, v_j\} \nsubseteq D(S)$.*

*3) $\sum_{v \in D(S)} \mathsf{score}(v)$ is maximized.*

*Intuitively, $D(S)$ is the set of at most $k$ results, such that no two results are similar with each other, and the total score of the results is maximized. We use $\mathsf{score}(D(S))$ to denote the total score of results in $D(S)$, i.e., $\mathsf{score}(D(S)) = \sum_{v \in D(S)} \mathsf{score}(v)$.*

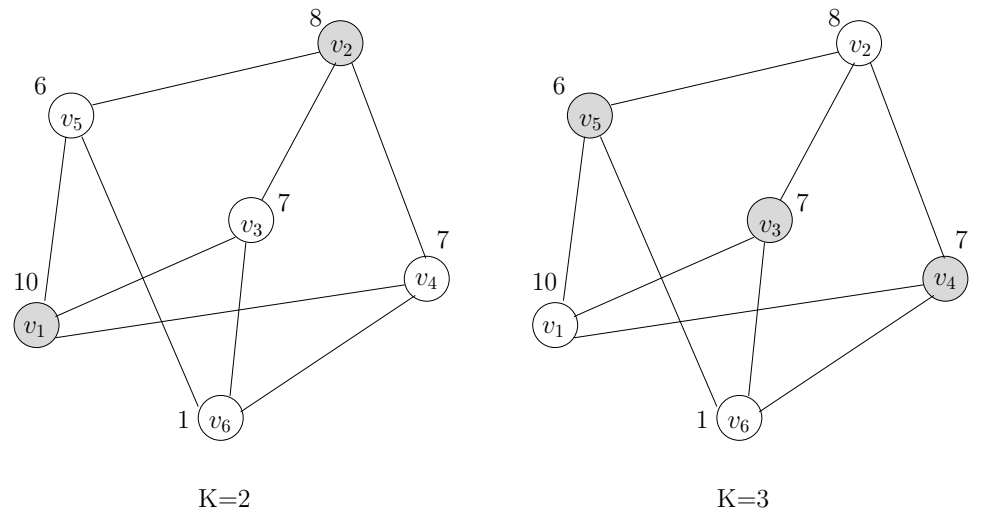

**Figure 1: A sample diversity graph**

In this paper, we are to find the diversified top-$k$ results. Our aim is to find a general approach, such that for any existing algorithm that returns the top-$k$ results of a certain problem, it can be easily changed to return the diversified top-$k$ results by applying our framework, in which the result set $S$ is not necessarily to be computed in advanced but grows incrementally with an early stop condition. We first give the definition of the diversity graph.

**Definition 2 (Diversity Graph)** *Given a list of results $S = \{v_1, v_2, \cdots\}$, the diversity graph of $S$, denoted as $G(S) = (V, E)$, is an undirected graph such that for any result $v \in R$, there is a corresponding node $v \in V$, and for any two results $v_i \in S$ and $v_j \in R$, there is an edge $(v_i, v_j) \in E$ iff $v_i \approx v_j$. We use $V(G(S))$ and $E(G(S))$ to denote the set of nodes and the set of edges in the diversity graph $G(S)$ respectively, and use $v.\mathsf{adj}(G(S))$ to denote the set of nodes that are adjacent to $v$ in $G(S)$. If the context is obvious, we use $v_i$ to denote both the result $v_i$ in $S$ and the node $v_i$ in $G(S)$, we use $G$ to denote $G(S)$, and we use $D$ to denote $D(S)$. Without loss of generality, we assume nodes in $G(S)$ are arranged in non-increasing order of their scores, i.e., for any $1 \leq i < j \leq |V(G(S))|$, $\mathsf{score}(v_i) \geq \mathsf{score}(v_j)$.*

The diversified top-$k$ results $D(S)$ can be equivalently defined as a subset of nodes in $G(S)$, that satisfy the three conditions.

1) $|D(S)| \leq k$.

2) $D(S)$ is an independent set of $G(S)$.

3) $\mathsf{score}(D(S))$ is maximized.

Here, an independent set of a graph is a set of nodes in a graph, where no two nodes are adjacent.

**Example 1** *Fig. 1 shows the diversity graph for 6 results $S = \{v_1, v_2, \cdots, v_6\}$. Suppose $k = 2$, the optimal solution $D(S)$ includes two points $v_1$ and $v_2$ with score $18$, as shown on the left part of Fig. 1. Suppose $k = 3$, the optimal solution $D(S)$ includes three points $v_3$, $v_4$ and $v_5$ with score $20$, as shown on the right part of Fig. 1.*

In the following, we first show the two existing frameworks to solve top-$k$ search problems, namely, the incremental top-$k$ framework and the bounding top-$k$ framework, which are most generally used in top-$k$ search algorithms. Then we show the framework of

**Algorithm 1** incremental($k$)

1: $S \leftarrow \emptyset$;
2: **for** i=1 **to** k **do**
3: $\quad v \leftarrow$ incremental-next();
4: $\quad$ **if** $v = \emptyset$ **then**
5: $\quad\quad$ **break**;
6: $\quad S \leftarrow S \bigcup \{v\}$;
7: **return** $S$;

**Algorithm 2** bounding($k$)

1: $S \leftarrow \emptyset$;
2: $\overline{\mathsf{unseen}} \leftarrow +\infty$;
3: **while** the $k$-th largest score of $S < \overline{\mathsf{unseen}}$ **do**
4: $\quad v \leftarrow$ bounding-next();
5: $\quad$ **if** $v = \emptyset$ **then**
6: $\quad\quad$ **break**;
7: $\quad S \leftarrow S \bigcup \{v\}$;
8: $\quad$ update $\overline{\mathsf{unseen}}$;
9: **return** top-$k$ results in $S$;

our approach to extend the two frameworks to handle diversified top-$k$ search.

# 3. TOP-$K$ SEARCH FRAMEWORKS

In the literature, the framework of most algorithms that find top-$k$ results falls into two categories, namely, the incremental top-$k$ framework and the bounding top-$k$ framework.

**Incremental Top-$k$:** In the incremental top-$k$ framework, results are generated one by one by calling a procedure incremental-next(), with non-increasing order of their scores. The algorithm stops after $k$ results are generated, and the $k$ results are the final top-$k$ results for the problem. The framework named incremental is shown in Algorithm 1. A lot of existing work fall into this category, e.g., finding top-$k$ shortest paths in graphs, finding top-$k$ steiner trees, communities and $r$-cliques in graphs, etc [16, 15, 20, 14]. A lot of works have been done to assume that the time complexity of each incremental-next() procedure to generate the next result with largest score is polynomial w.r.t. the size of the input only.

**Bounding Top-$k$:** In the bounding top-$k$ framework, results are generated one by one by calling a procedure bounding-next(), but not necessarily with non-increasing order of their scores. A bound $\overline{\mathsf{unseen}}$ is defined to be the upper bound of the scores for the unseen results. After each result is generated by bounding-next(), $\overline{\mathsf{unseen}}$ is also updated to be a possibly smaller value. The algorithm stops when the $k$-th largest score of all generated results is no smaller than the upper bound for the unseen results $\overline{\mathsf{unseen}}$. The framework named bounding is shown in Algorithm 2. The threshold algorithm that is generally used to return top-$k$ results falls into this category [7, 9]. Other works that fall into this category include [12, 17].

# 4. DIVERSIFIED TOP-$K$ SEARCH

In this section, we show how to extend the incremental top-$k$ framework incremental and bounding top-$k$ framework bounding to handle diversified top-$k$ search. We mainly focus on two tasks. First, a new early stop conditions is needed. Second, an algorithm that finds the diversified top-$k$ results for the current generated result set is needed. For the early stop condition, in the original algorithm, the stop condition for incremental is simply $|S| = k$ and the stop condition for bounding is the current $k$-th largest score $\leq \overline{\mathsf{unseen}}$. Obviously, both of them cannot be applied to handle diversified top-$k$ search. Consider an extreme case, when the algorithm stops using the original stop condition, it is possible that all the results generated are similar to each other. Thus the current diversified top-$k$ results only contain 1 result with the largest score. It is not the optimal solution because it is possible that an unseen result is not similar to the current one. Here, $D(S)$ computed for the current generated result set $S$ can be used to check the new stop condition, and if the new stop condition is satisfied, $D(S)$ is the optimal solution for the diversified top-$k$ search.

**Algorithm 3** div-search($k$)

1: $S \leftarrow \emptyset$; $D(S) \leftarrow \emptyset$;
2: **while** sufficient() **do**
3: $\quad$ the code to update $S$ (and $\overline{\mathsf{unseen}}$);
4: $\quad$ **if** necessary() **then**
5: $\quad\quad D(S) \leftarrow$ div-search-current($G(S), k$);
6: **return** $D(S)$;

We extend both incremental and bounding using the same framework, which is shown in Algorithm 3, by adding three new functions, a new sufficient stop condition sufficient(), a new necessary stop condition necessary() and an algorithm div-search-current() to search the diversified top-$k$ results on the current generated result set. The algorithm executes the code of the original top-$k$ algorithm to update $S$ and stops when sufficient() is satisfied. For incremental, the code is line 3-6 in algorithm 1, and for bounding, the code is line 4-8 in algorithm 2. After updating $S$, we construct the diversity graph $G(S)$ on $S$ based on the similarity function sim() for any given two results. If the necessary stop condition is satisfied, we find the diversified top-$k$ results for the current result set $S$ using div-search-current(). The necessary stop condition is used to reduce the number of calling div-search-current(), because div-search-current() is a costly work. In the following, we will introduce the sufficient stop condition, the necessary stop condition, and the search algorithm for current set.

**Sufficient Stop Condition:** Given the current result set $S$, we need to calculate an upper bound $\overline{\mathsf{best}}(S)$ for the possible optimal solutions considering both the current result set $S$ and the unseen results. Let $D_i(S)$ be the best diversified results of $S$ with exactly $i$ elements for $1 \leq i \leq k$, i.e., $D_i(S)$ is a subset of nodes in $V(G(S))$, that satisfies the following three conditions.
1) $|D_i(S)| = k$.
2) $D_i(S)$ is an independent set of $G(S)$.
3) $\mathsf{score}(D_i(S))$ is maximized.

**Lemma 1** *Given $D_i(S)$ for $1 \leq i \leq k$ and the score upper bound of all the unseen results $\overline{u}$. The upper bound $\overline{\mathsf{best}}(S)$ can be calculated as follows.*

$$\overline{\mathsf{best}}(S) = \max_{1 \leq i \leq k} \{\mathsf{score}(D_i(S)) + (k - i) \times \overline{u}\} \tag{1}$$

*where $\overline{u}$ is the score of the last generated result $v$,* $\mathsf{score}(v)$*, for* incremental *and is the upper bound of the unseen results,* $\overline{\mathsf{unseen}}$*, for* bounding.

**Proof Sketch:** Suppose the final optimal solution is $O$, then we can divide $O$ into two parts, $O = O_1 \bigcup O_2$, where $O_1$ is the set of generated results, and $O_2$ is the set of unseen results. Suppose $O_1$ has $n_1$ elements and $O_2$ has $n_2$ elements. We have $n_1 + n_2 \leq k$. Since $O_1$ is the set of generated results, we have (1) $\mathsf{score}(O_1) \leq \mathsf{score}(D_{n_1}(S))$, since $D_{n_1}(S)$ is the optimal solution with $n_1$ elements. We also have (2) $\mathsf{score}(O_2) \leq n_2 \times \overline{u} \leq (k - n_1) \times \overline{u}$, since $\overline{(}u)$ is the score upper bound for all unseen results. Combine (1) and (2), we have $\mathsf{score}(O) = \mathsf{score}(O_1) + \mathsf{score}(O_2) \leq score(D_{n_1}(S)) + (k - n_1) \times \overline{u} \leq \max_{1 \leq i \leq k}\{\mathsf{score}(D_i(S)) + (k-i) \times \overline{u}\} = \overline{\mathsf{best}}(S)$. $\overline{\mathsf{best}}(S)$ is an upper bound for the optimal solution. $\square$

Having the score upper bound $\overline{\mathsf{best}}(S)$ for the optimal solution, the sufficient stop condition for div-search can be defined as follows.

$$\mathsf{score}(D(S)) \geq \overline{\mathsf{best}}(S) \quad (2)$$

The following lemma shows that, after every iteration, div-search moves towards the sufficient stop condition.

**Lemma 2** *For any* $S' \subseteq S$, $\mathsf{best}(S') \leq \mathsf{best}(S)$ *and* $\overline{\mathsf{best}}(S') \geq \overline{\mathsf{best}}(S)$.

**Proof Sketch:** Since $S' \subseteq S$, the best solution on $S'$ is a feasible solution on $S$, thus $\mathsf{best}(S') \leq \mathsf{best}(S)$. Comparing to $\overline{\mathsf{best}}(S')$, $\overline{\mathsf{best}}(S)$ is calculated by changing some upper bounds $\overline{u'}$ when calculating $\mathsf{best}(S')$ into the real scores no larger than $\overline{u'}$ and changing the other unseen upper bounds from $\overline{u'}$ to $\overline{u}$, where $\overline{u} \leq \overline{u'}$ is assumed by the original algorithm. Thus $\overline{\mathsf{best}}(S') \geq \overline{\mathsf{best}}(S)$. $\square$

**Necessary Stop Condition:** We discuss the necessary stop condition for div-search. The necessary stop condition is used as follows. In each iteration, before invoking div-search-current(), if the necessary stop condition is not satisfied, then div-search-current() is not necessarily to be invoked in this iteration.

**Lemma 3** *For* div-search*, if it can stop in a certain iteration, one of the following conditions should be satisfied before invoking the procedure* div-search-current()*:*
*1) The last generated result* $v = \emptyset$.
*2)* $|S| \geq |S'| + k - \max\{i | 1 \leq i \leq k, D_i(S') \neq \emptyset\}$ *and the k-th largest score in* $S \geq \overline{u}$.
*Here* $S'$ *is the set of results when the last* div-search-current() *is invoked or* $\emptyset$ *if* div-search-current() *is never invoked.*

**Proof Sketch:** The first condition is trivial. Now suppose $v \neq \emptyset$. For the second condition, when the $k$-th largest score in $S < \overline{u}$, it is possible that a new result can be added that updates the $k$-th largest score, and thus improves the current best solution. Now we discuss $|S| \geq |S'| + k - \max\{i | 1 \leq i \leq k, D_i(S') \neq \emptyset\}$. $\max\{i | 1 \leq i \leq k, D_i(S') \neq \emptyset\}$ is the size of the maximum independent set for $G(S')$ if it is smaller than $k$, and $k - \max\{i | 1 \leq i \leq k, D_i(S') \neq \emptyset\}$ is the minimum number of nodes needed to be added in order to generate a result of size $k$. If such a result does not exist, we cannot stop because we can always add some unseen nodes to any existing solution with a size smaller than $k$ to make the score larger. As a result, we should add at least $k - \max\{i | 1 \leq i \leq k, D_i(S') \neq \emptyset\}$ nodes into $S'$. $\square$

**Searching Current Set:** The most important operation in our framework is the the algorithm div-search-current() to search the diversified top-$k$ results for the current result set $S$. We first show the difficulties of the problems in this section and give three algorithms, namely div-astar, div-dp, and div-cut on div-search-current() in the next three sections respectively.

The following lemma shows that finding the diversified top-$k$ results is an NP-Hard problem.

**Lemma 4** *Finding* $D(S)$ *on* $G(S)$ *is an NP-Hard problem.*

**Proof Sketch:** We consider a special case of the problem, where $\mathsf{score}(v) = 1$ for all $v \in V(G(S))$, and $k = |V(G(S))|$. In such a case, finding $D_k(R)$ on $G(S)$ is equivalent to finding the maximum independent set on graph $G(S)$, which is an NP-Hard problem. Thus, the original problem is an NP-Hard problem.

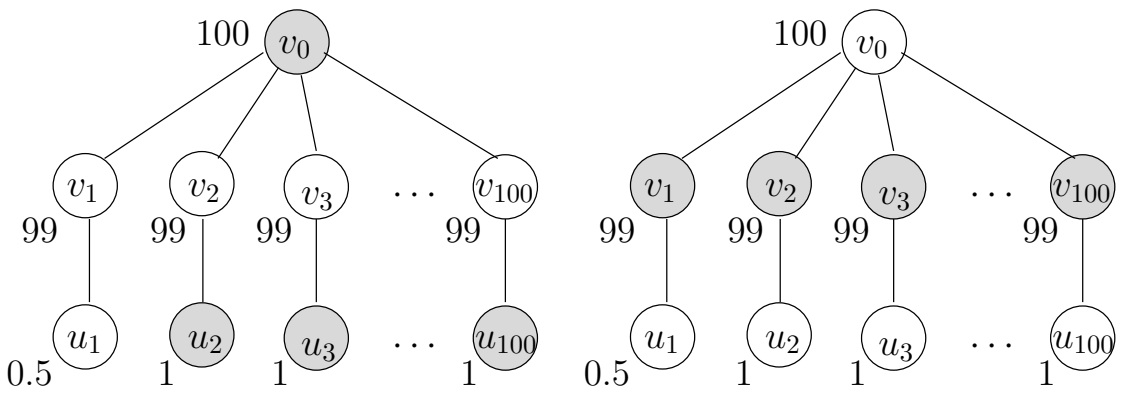

(a) The Greedy Solution (b) The Optimal Solution

**Figure 2: The greedy algorithm**

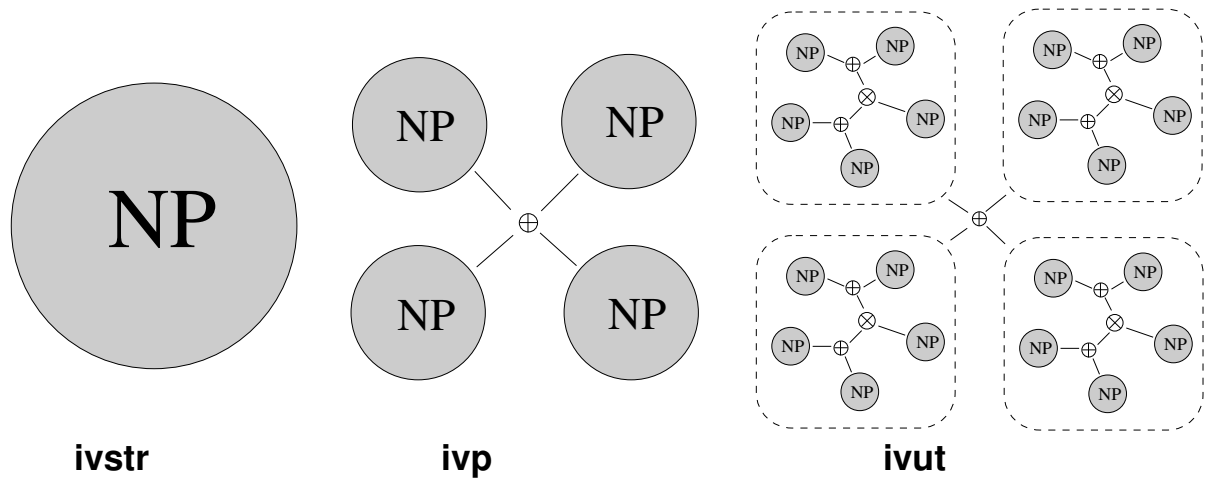

**Figure 3: Overview of three algorithms**

**Greedy is Not Good:** Given $G(S)$ and $k$, a simple greedy algorithm to find $D(S)$ works as follows. It processes in iterations. In each iteration, the node $v$ with the maximum score is selected and put into $D(S)$. After that, all the nodes that are adjacent to $v$ in $G(S)$ is removed from $G(S)$. The process stops when $G(S)$ is empty or $D(S)$ contains $k$ results.

The quality of the greedy algorithm can be arbitrarily bad. The approximation ratio for the greedy algorithm is not bounded by a constant factor. Even for its special case, the maximum independent set problem is known to be hard to approximate in the literature. We give an example. Fig. 2 shows a diversity graph with 201 nodes and 200 edges. Suppose $k = 100$. Using the greedy algorithm, the solution is shown in Fig. 2(a), where the selected results are marked gray. The score of the greedy solution is 199. The optimal solution for the problem is shown in Fig. 2(b). The score of the optimal solution is $9,900$, which is nearly 50 times of the score of the greedy solution.

In the following, we propose to find the optimal solution of $D(S)$. We propose three algorithms, namely, div-astar, div-dp, and div-cut. div-astar searches the whole space $S$ using the A$^*$ based heuristics by designing an upper bound function astar-bound(). Based on the NP-Hardness of the problem, div-astar can hardly handle problems with large diversity graph $G$. In our second div-dp algorithm, we decompose $G$ into connected components. The size of each component can be much smaller than the original graph $G$, and is searched independently using div-astar. We combine the components using an efficient operation $\oplus$ based on dynamic programming. In our third div-cut algorithm, we further decompose each connected component into subgraphs, where subgraphs are connected through a set of cut points. Each subgraph is searched independently for at most 4 times under different conditions. We combine the components using two efficient operations $\oplus$ and $\otimes$. The general ideas of the three algorithms are illustrated in Fig. 3.

# 5. AN A$^*$ BASED APPROACH

As discussed in Section 4, div-search-current$(G(S), k)$ should return the optimal solution $D_i(S)$ for $1 \leq i \leq k$ in order to find the early stop condition. For simplicity, we use $D$ to denote the set of solutions, and we use $D.\mathsf{solution}_i$ to denote the optimal solution

**Algorithm 4** div-astar$(G, k)$

**Input**: The diversity graph $G$, the top-$k$ value.
**Output**: Search result $D$.

1: $\mathcal{H} \leftarrow \emptyset$; $D \leftarrow \emptyset$;
2: $\mathcal{H}$.push$((\emptyset, 0, 0, 0))$;
3: **for** $k' = k$ **down to** 1 **do**
4: astar-search$(G, \mathcal{H}, D, k')$;
5: **for all** $e \in \mathcal{H}$ **do**
6: $e.\overline{\text{bound}} \leftarrow$ astar-bound$(G, e, k')$;
7: update $e$ in $\mathcal{H}$;
8: **return** $D$;

9: **procedure** astar-search$(G, \mathcal{H}, D, k')$
10: **while** $\mathcal{H} \neq \emptyset$ **and** $\mathcal{H}.\text{top}.\overline{\text{bound}} > \max_{i \leq k'}\{D.\text{score}_i\}$ **do**
11: $e \leftarrow \mathcal{H}$.pop();
12: **for** $i = e.\text{pos} + 1$ **to** $|V(G)|$ **do**
13: **if** $v_i.\text{adj}(G) \bigcap e.\text{solution} = \emptyset$ **then**
14: $e' \leftarrow (e.\text{solution} \bigcup \{v_i\}, i, e.\text{score} + \text{score}(v_i), 0)$;
15: $e'.\overline{\text{bound}} \leftarrow$ astar-bound$(G, e', k')$;
16: $\mathcal{H}$.push$(e')$;
17: update $D$ using $e'$.solution;

18: **procedure** astar-bound$(G, e, k')$
19: $p \leftarrow |e.\text{solution}|$; $i \leftarrow e.\text{pos} + 1$;
20: $\overline{\text{bound}} \leftarrow e.\text{score}$;
21: **while** $p < k'$ **and** $i < |V(G)|$ **do**
22: **if** $v_i.\text{adj}(G) \bigcap e.\text{solution} = \emptyset$ **then**
23: $\overline{\text{bound}} \leftarrow \overline{\text{bound}} + \text{score}(v_i)$;
24: $p \leftarrow p + 1$;
25: $i \leftarrow i + 1$;
26: **return** $\overline{\text{bound}}$;

with $i$ results $D_i(S)$, and use $D.\text{score}_i$ to denote the score for the optimal solution score$(D_i(S))$.

Our first algorithm is an A* based algorithm. The algorithm is shown in Algorithm 4. We define a max heap $\mathcal{H}$ to store the entries in the A* search. Each entry $e \in \mathcal{H}$ is with the form $e = (\text{solution}, \text{pos}, \text{score}, \overline{\text{bound}})$. Each entry $e$ is ranked in $\mathcal{H}$ according to $e.\overline{\text{bound}}$, which is the estimated upper bound of the solution if we further expand it in the A* search. $e$.solution is the partial solution searched and $e$.pos is the position of the last searched node in $e$.solution. $e$.score is the score of the partial solution, i.e., $e.\text{score} = \text{score}(e.\text{solution})$. The algorithm should return $D.\text{solution}_i$ for all $1 \leq i \leq k$. Suppose we have an A* algorithm that finds the optimal solution for a certain $D.\text{solution}_i$, the algorithm should be invoked $k$ times to find the $k$ solutions, which is costly. We show that after searching $D.\text{solution}_i$ for a certain $i$, the partial solutions in $\mathcal{H}$ can be reused when searching $D.\text{solution}_j$ for $j < i$. In the following, we first discuss the estimated upper bound for partial solutions. Then we discuss the A* algorithm to find the optimal solution $D.\text{solution}_i$ for a certain $i$. At last, we discuss how the partial solutions in $\mathcal{H}$ can be reused to find the optimal solutions $D.\text{solution}_i$ for all $1 \leq i \leq k$.

**Upper Bound Estimation:** Given a partial solution $e$, for a certain $k'$, we show how to estimate the score upper bound if we expand the partial solution to be a solution of at most $k'$ elements. The algorithm astar-bound is shown in Algorithm 4, line 18-26. The newly added nodes should at least satisfy the following two conditions: 1) they can not be one of $e$.solution, and 2) they are not adjacent to any node in $e$.solution. Under such conditions, we can just add the set of nodes with largest scores, and after adding the nodes, the total number of nodes is no larger than $k'$. In order to satisfy condition 1), we visit nodes in $G$ from the position $e.\text{pos} + 1$ (line 19). Since nodes in $G$ are sorted in the non-increasing order of their scores, we add nodes one by one until the size $p$ reaches $k'$. For each node added, condition 2) can be checked using $v_i.\text{adj}(G) \bigcap e.\text{solution} = \emptyset$ (line 22).

**Lemma 5** astar-bound$(G, e, k')$ *finds the score upper bound for the partial solution* $e$.solution *to be expanded to a solution of at most* $k'$ *elements.*

**Proof Sketch:** Suppose we have removed all the nodes from $G$ that are adjacent to at least one node in $e$.solution, then the function astar-bound$(G, e, k')$ calculates the upper bound by expanding $e$.solution using the set of nodes after position $e$.pos in $G$ with largest scores. The optimal solution that $e$.solution can be expanded also selects the expanded nodes from the set of nodes after position $e$.pos but it may not select all with the largest scores since some of them may be adjacent to each other. Thus the optimal solution can not be larger than astar-bound$(G, e, k')$. As a result, astar-bound$(G, e, k')$ is a score upper bound for all expansions of $e$.solution. □

**A* Search for a Certain $k$:** To find the optimal solution for a certain $k = k'$, the A* search algorithm astar-search is shown in Algorithm 4, line 9-17. It runs in iterations. In each iteration, the partial solution $e$ with the largest estimated upper bound is popped out from $\mathcal{H}$ (line 11). $e$ can then be expanded to new partial solutions by adding a new node into $e$.solution. The nodes are added from position $e.\text{pos} + 1$ in $G$ since all nodes before the position has been processed (line 12). The newly added node $v_i$ should not be adjacent to one of $e$.solution(line 13), and after adding the new node, the upper bound of the new partial solution should be updated using astar-bound(), and the new partial solution should be pushed into $\mathcal{H}$ for further expansion (line 14-16). In line 17, suppose the new partial solution $e'$ has $j$ elements, the new partial solution is considered as a solution of size $j$, and used to update the current best solution $D.\text{solution}_j$, if $D.\text{score}_j < e'.\text{score}$. The iteration stops if either $\mathcal{H}$ is $\emptyset$ or the largest upper bound in $\mathcal{H}$ is no larger than the current best score $\max_{i \leq k'}\{D.\text{score}_i\}$ (line 10).

**Reusing Partial Solutions:** In Algorithm 4, line 1-8 show how to share the same $\mathcal{H}$ to compute $D.\text{solution}_{k'}$ for $1 \leq k' \leq k$, without constructing $\mathcal{H}$ from scratch each time $k'$ changes. It processes with decreasing order of $k'$ (line 3). After processing $k'$, the partial solutions in $\mathcal{H}$ can be reused when processing $k' - 1$, in order to save computational cost. If we simply keep the current entries in $\mathcal{H}$, they cannot be used directly to process $k' - 1$. It is because the upper bounds for each partial solution are calculated by expanding to a solution of size $k'$, which is not the upper bounds for a solution of size $k' - 1$. In order to reuse the partial solutions in $\mathcal{H}$, we need to recalculate the upper bounds for all partial solutions in $\mathcal{H}$ using $k' - 1$ and update the new positions for elements in $\mathcal{H}$ (line 5-7). The following lemma shows the correctness of the approach.

**Lemma 6** *The partial solutions in* $\mathcal{H}$ *for calculating* $D.\text{solution}_i$ *can be reused when calculating* $D.\text{solution}_{i-1}$.

**Proof Sketch:** Consider the possibly expanded node $e$ in $\mathcal{H}$ such that $e.\text{solution} = D.\text{solution}_{i-1}$. There is a unique path from the root of $\mathcal{H}$ to $e$. (1) Suppose $e$ is not removed from $\mathcal{H}$ currently, then there exists a unique ancestor of $e$ in the current $\mathcal{H}$. Since the upper bounds have been updated and $e$ is the optimal solution $D.\text{solution}_{i-1}$, $e$ can be expanded when calculating $D.\text{solution}_{i-1}$. (2) Suppose $e$ has been removed from $\mathcal{H}$ currently, then $e$ has been used to update $D.\text{solution}_{i-1}$ after removal. Since the upper bounds for all entries in $\mathcal{H}$ have been updated and $e$ is the optimal solution $D.\mathsf{solution}_{i-1}$, the process can stop directly before popping any entry from $\mathcal{H}$. From (1) and (2), we conclude that after reusing the partial solution, $D.\mathsf{solution}_{i-1}$ can be calculated. □

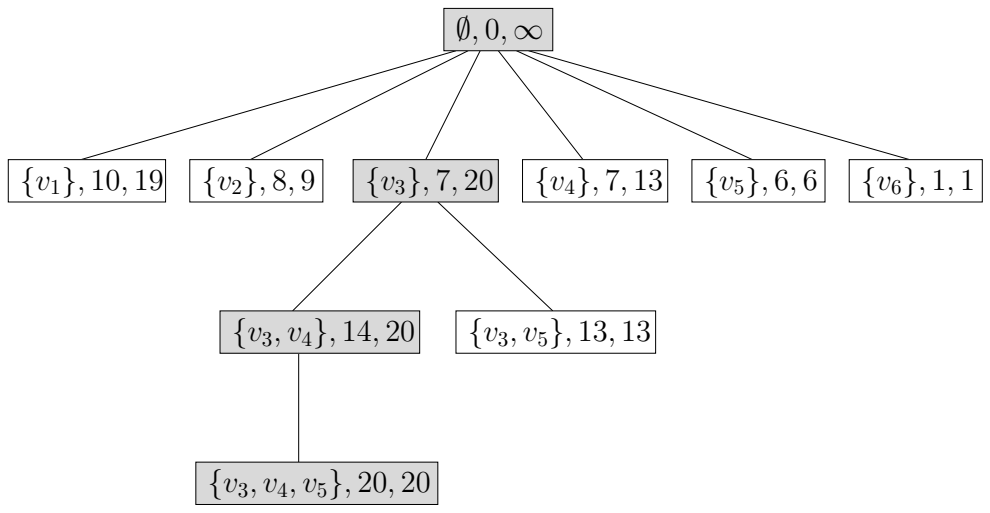

**Figure 4: Finding diversified top-3 results**

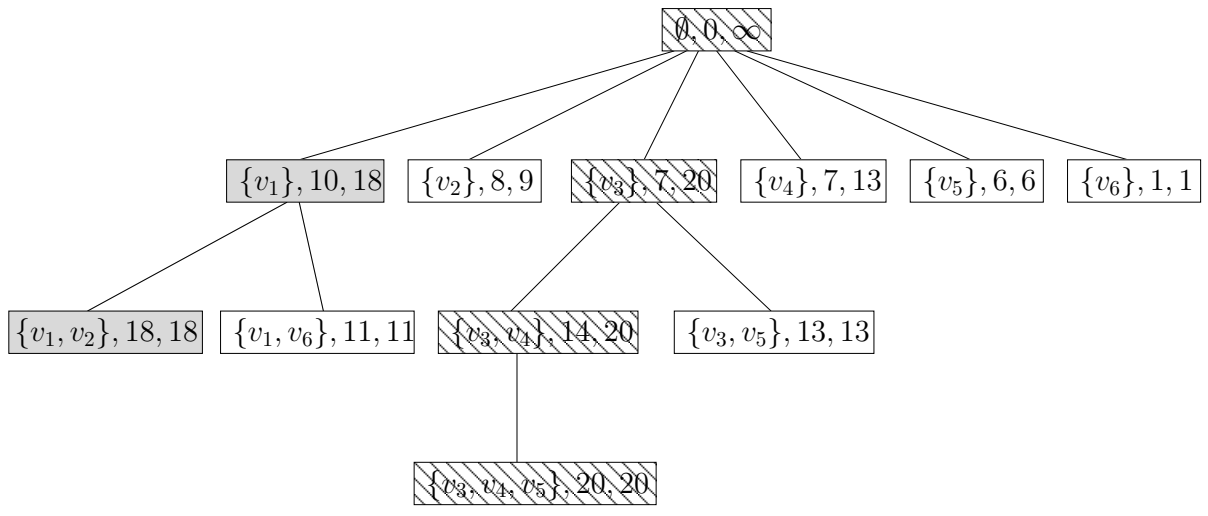

**Figure 5: Finding diversified top-2 results**

**Example 2** *Consider the diversity graph shown in Fig. 1. Suppose $k = 3$. The process for the* div-astar *search is shown in Fig. 4. First, entry $(\emptyset, 0, \infty)$ is popped from the heap $\mathcal{H}$, and 6 new entries are pushed. The result $\{v_1\}$ is with upper bound 19 because after selecting $v_1$, nodes $v_3$, $v_5$ and $v_4$ that are adjacent to $v_1$ should be excluded when calculating the upper bound of $\{v_1\}$. The only nodes left are $v_2$ and $v_6$ with scores 8 and 1 respectively. Thus the upper bound for $\{v_1\}$ is* $\mathsf{score}(\{v_1\}) + 8 + 1 = 19$. *The one with the highest score is $(\{v_3\}, 7, 20)$, and thus it is popped from $\mathcal{H}$ in the next iteration and generates two new entries. Accordingly, $(\{v_3, v_4\}, 14, 20)$ and $(\{v_3, v_4, v_5\}, 20, 20)$ are popped from $\mathcal{H}$ in order. At this moment, $D.\mathsf{score}_1 = 10$, $D.\mathsf{score}_2 = 14$, and $D.\mathsf{score}_3 = 20$. The stop condition is satisfied, and $D.\mathsf{solution}_3$ is the optimal diversified top-3 results. Consider now we compute $D.\mathsf{solution}_2$ since it is not the optimal solution currently. We do not need to reconstruct $\mathcal{H}$ from scratch. We update all upper bounds for entries that exists on the current $\mathcal{H}$. In this example, entry $(\{v_1\}, 10, 19)$ is updated to be $(\{v_1\}, 10, 18)$ as shown in Fig. 5. Continue the iteration, $(\{v_1\}, 10, 18)$ and $(\{v_1, v_2\}, 18, 18)$ are popped from $\mathcal{H}$ in order. At this moment, $D.\mathsf{score}_2$ is updated to be 18 and the stop condition is satisfied. Thus 18 is the best score for $k = 2$.*

## 6. A DP BASED APPROACH

The div-astar algorithm is not suitable to handle large diversity graph $G$ since the search space for div-astar increases exponentially with respect to the size of $G$ and $k$. In order to reduce the size of the diversity graph $G$ used for div-astar search. In this section, we decompose $G$ into a set of disconnected components. We show that we only need to process each disconnected component separately using div-astar, and the solution for each disconnected component can be combined efficiently to the solution of the whole graph $G$ using dynamic programming. Before introducing the algorithm, we first introduce two operators, $\oplus$ and $\otimes$.

**Algorithm 5** operator $\oplus(D', D'')$

**Input**: Search results $D'$ and $D''$ for subgraphs.
**Output**: Search result $D = D' \oplus D''$.

```
1: D ← ∅;
2: for i = 1 to k do
3:     D.score_i ← 0; D.solution_i ← ∅;
4:     for j = 0 to i do
5:         if D'.solution_j ≠ ∅ or j = 0 then
6:             if D''.solution_{i−j} ≠ ∅ or i = j then
7:                 if D'.score_j + D''.score_{i−j} > D.score_i then
8:                     D.score_i ← D'.score_j + D''.score_{i−j};
9:                     D.solution_i ← D'.solution_j ⋃ D''.solution_{i−j};
10: return D;
```

**Algorithm 6** operator $\otimes(D', D'')$

**Input**: Search results $D'$ and $D''$ for subgraphs.
**Output**: Search result $D = D' \otimes D''$.

```
1: D ← ∅;
2: for i = 1 to k do
3:     D.score_i ← 0; D.solution_i ← ∅;
4:     if D'.score_i > D''.score_i then
5:         D.score_i ← D'.score_i;
6:         D.solution_i ← D'.solution_i;
7:     else
8:         D.score_i ← D''.score_i;
9:         D.solution_i ← D''.solution_i;
10: return D;
```

**The $\oplus$ Operator:** The $\oplus$ operator has two operands, search result $D'$ and search result $D''$. For $1 \leq i \leq k$, $D.\mathsf{solution}_i$ is the solution of size $i$ with the largest score by combining some nodes in $D'$ and other nodes in $D''$. The algorithm to compute $D = D' \oplus D''$ is shown in Algorithm 5 using dynamic programming. It calculates $D.\mathsf{solution}_i$ one by one for $1 \leq i \leq k$ (line 2). For a certain $i$, we try to select $j$ nodes from $D'$ and the left $i - j$ nodes from $D''$ for all $0 \leq j \leq i$ (line 4). For a certain $j$, it can generate a feasible solution from $D'$ and $D''$ if the two conditions are satisfied: 1) $D'.\mathsf{solution}_j \neq \emptyset$ or $j = 0$, and 2) $D''.\mathsf{solution}_{i-j} \neq \emptyset$ or $i - j = 0$ (line 5-6). $D.\mathsf{solution}_i$ is the one that results in the largest total score (line 7-9). The time complexity for Algorithm 5 is $O(k^2)$. The $\oplus$ operator is suitable to operate on two search results that are generated from two disjoint subgraphs respectively. The $\oplus$ operator has the following two properties.
(Commutative law) $D \oplus D' = D' \oplus D$.
(Associative law) $(D \oplus D') \oplus D'' = D \oplus (D' \oplus D'')$.

**The $\otimes$ Operator:** Similar to the $\oplus$ operator, the $\otimes$ operator is operated on two operands, search result $D'$ and search result $D''$. For $1 \leq i \leq k$, $D.\mathsf{solution}_i$ is the solution of size $i$ that are the best of $D'.\mathsf{solution}_i$ and $D''.\mathsf{solution}_i$. The algorithm to compute $D = D' \otimes D''$ is shown in Algorithm 6. It calculates $D.\mathsf{solution}_i$ one by one for $1 \leq i \leq k$ (line 2). For a certain $i$, $D.\mathsf{solution}_i$ is set to be $D'.\mathsf{solution}_i$ if $D'.\mathsf{score}_i > D''.\mathsf{score}_i$, and is set to be $D''.\mathsf{solution}_i$ otherwise (line 4-9). The time complexity for Algorithm 6 is $O(k)$. The $\otimes$ operator is suitable to operate on two search results that are generated from the same subgraph. The $\otimes$ operator will be used and discussed in the next section. The $\otimes$ operator has the following two properties.
(Commutative law) $D \otimes D' = D' \otimes D$.
(Associative law) $(D \otimes D') \otimes D'' = D \otimes (D' \otimes D'')$.

**The Overall Approach:** The overall approach to compute $D$ is shown in Algorithm 7. We find the set of disconnected components

**Algorithm 7** div-dp$(G, k)$

**Input**: The diversity graph $G$, the top-$k$ value.
**Output**: Search result $D$.

1: $D \leftarrow \emptyset$;
2: let $C = \{G_1, G_2, \cdots\}$ be the set of connected components of $G$;
3: **for all** $G_i \in C$ **do**
4: $\quad D' \leftarrow$ div-astar$(G_i, k)$;
5: $\quad D \leftarrow D \oplus D'$;
6: **return** $D$;

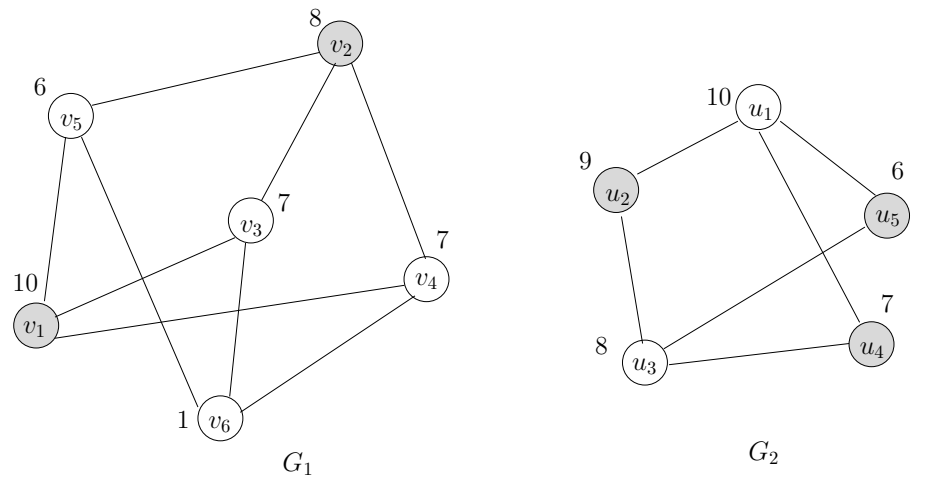

**Figure 6: A sample diversity graph**

$C$ of $G$ (line 2). We then process each $G_i \in C$ individually using div-astar search (line 4). With the commutative law and associative low of operator, the search results for subgraphs in $C$ can be combined with $D$ in an arbitrary order using operator $\oplus$ (line 5).

**Example 3** *Consider the diversity graph G shown in Fig. 6 that contains two connected components $G_1$ and $G_2$. Suppose $k = 5$, and suppose the results $D_1$ of $G_1$ and $D_2$ of $G_2$ have been computed using* div-astar *algorithm separately. We now combine $D_1$ and $D_2$ to compute the result for D for G. Suppose we now compute $D$.solution$_5$, using the $\oplus$ operator shown in Algorithm 5, if we select 1 node from $G_1$ and select 4 nodes from $G_2$, we got a score 0 since $D_2$.solution$_4 = \emptyset$. If we select 2 nodes from $G_1$ and select 3 nodes from $G_2$, we got a score 40, and if we select 3 nodes from $G_1$ and 2 nodes from $G_2$ we got a score 38. We search all the possible combinations, and select the best of them, which is 40, with 2 nodes from $G_1$ and 3 nodes from $G_2$, as the best solution $D$.solution$_5$. The operation $\oplus$ to combine $D_1$ and $D_2$ is shown in Fig. 7.*

# 7. A CUT POINT BASED APPROACH

The dynamic programming based approach divides the diversity graph $G$ into components and each component can be searched separately. When one of the components is large, it is still a costly work to search the single component. Consider a certain connected component $G_i$, although it is connected, it may contain several subgraphs that are loosely connected, i.e., connected through a set of cut points, where a cut point of a graph $G_i$ is a single node $v \in V(G_i)$ such that $G_i$ is disconnected if removing $v$ from $G_i$. In this section, we show that the subgraphs connected through some of the cut points can be considered separately by applying div-astar search at most 4 times under different assumptions, and their search results can be combined using a series of $\oplus$ and $\otimes$ operations.

**The Cut Point Tree (**cptree**):** Given a connected graph $G$, the cut point tree (cptree) of $G$ is a tree formed by a subset of cut points of $G$. Each node $o$ of cptree is with the form $o =$ ($o$.cut-point, $o$.entry-graph, $o$.left-graph, $o$.subnodes, $o$.result). $o$.cut-point is the corresponding cut point representing the node. $o$.entry-graph is the subgraph of $G$ that connects $o$.cut-point and the cut-point of the farther node of $o$ on cptree. If there are more than one such graphs, then $o$.entry-graph is a disconnected graph that contains

| solution | score |
|---|---|
| 0: $\emptyset$ | 0 |
| 1: $\{v_1\}$ | 10 |
| 2: $\{v_1, v_2\}$ | 18 |
| 3: $\{v_3, v_4, v_5\}$ | 20 |
| 4: $\emptyset$ | 0 |
| 5: $\emptyset$ | 0 |

$\oplus$

| solution | score |
|---|---|
| 0: $\emptyset$ | 0 |
| 1: $\{u_1\}$ | 10 |
| 2: $\{u_1, u_3\}$ | 18 |
| 3: $\{u_2, u_4, u_5\}$ | 22 |
| 4: $\emptyset$ | 0 |
| 5: $\emptyset$ | 0 |

$=$

| solution | score |
|---|---|
| 0: $\emptyset$ | 0 |
| 1: $\{v_1\}$ | 10 |
| 2: $\{v_1, u_1\}$ | 20 |
| 3: $\{v_1, u_1, u_3\}$ | 28 |
| 4: $\{v_1, v_2, u_1, u_3\}$ | 36 |
| 5: $\{v_1, v_2, u_2, u_4, u_5\}$ | 40 |

**Figure 7: Dynamic programming using the $\oplus$ operator**

**Algorithm 8** div-cut$(G, k)$

**Input**: The diversity graph $G$, the top-$k$ value.
**Output**: Search result $D$.

1: $D \leftarrow \emptyset$;
2: let $C = \{G_1, G_2, \cdots\}$ be the set of connected components of $G$;
3: **for all** $G_i \in C$ **do**
4: $\quad$ cp-compress$(G_i)$;
5: $\quad$ **if** cut-points$(G_i) = \emptyset$ **then**
6: $\quad\quad D' \leftarrow$ div-astar$(G_i, k)$;
7: $\quad$ **else**
8: $\quad\quad o \leftarrow$ cptree-construct$(G_i,$ cut-points$(G_i))$;
9: $\quad\quad$ cp-search$(G_i, o, k)$;
10: $\quad\quad D' \leftarrow o.$result$_0 \otimes o.$result$_1$;
11: $\quad D \leftarrow D \oplus D'$;
12: **return** $D$;

all of them. It is possible that $o$.entry-graph $= \emptyset$. $o$.left-graph is the graph that does not contain any cut point after removing $o$.cut-point from $G$. If there are more than one such graphs, then $o$.left-graph is a disconnected graph that contains all of them. It is possible that $o$.left-graph $= \emptyset$. $o$.subnodes is the set of subnodes for $o$ in the cptree. $o$.result contain two results, $o$.result$_0$ and $o$.result$_1$. $o$.result$_0$ is the search result for the subtree rooted at $o$ such that $o$.cut-point is excluded, and $o$.result$_1$ is the search result for the subtree rooted at $o$ such that $o$.cut-point is included. We use $o$.cut-point to denote $o$ if the context is obvious.

**Graph Compression:** In order to increase the number of cut points in a graph and thus reduce the size of the sub-components after removing some of the cut points. We study how to compress a graph $G$. By compression, we mean some nodes can be removed from the graph if the final solution $D$ on $G$ is not influenced. The following lemma shows how to compress a graph $G$.

**Lemma 7** *Given the diversity graph G, a node $v_i$ can be removed from G if there exists a node $v_j$ that satisfies the following three conditions.*
*1) $v_j \in v_i.adj(G)$.*
*2) $\mathsf{score}(v_j) \geq \mathsf{score}(v_i)$.*
*3) $v_j.adj(G) \bigcup \{v_j\} \subseteq v_i.adj(G) \bigcup \{v_i\}$.*
*After removing $v_i$, the optimal solution on the new graph is the same with the optimal solution on the original graph.*

**Proof Sketch:** We prove that for any solution $V$ that contains $v_i$, we can get a solution by replacing $v_i$ with $v_j$ and the score is not decreased. First, we prove after replacing $v_i$ with $v_j$, the solution is still a feasible solution. Since $v_i$ and $v_j$ are adjacent (the first condition), $v_j$ can not be contained in the original solution. Since each node that $v_j$ connects is connected to $v_i$ (the third condition), and there are no nodes in $V$ that are adjacent to $v_i$, after replacing $v_i$ with $v_j$ in $V$, there are still no nodes in $V$ that are adjacent to $v_j$. Thus, the new solution is still an feasible solution. Since $\mathsf{score}(v_j) \geq \mathsf{score}(v_i)$ (the second condition), the score of the new solution is no smaller than the score of the original solution $V$. $\square$

**Algorithm 9** cptree-construct($G$, cut-points)

**Input**: Graph $G$, a set of cut points cut-points.
**Output**: The root cpnode $o$ of the cptree.

1: cpnode $o \leftarrow \emptyset$;
2: $v \leftarrow$ a node in cut-points with smallest entry-graph($v$);
3: $o$.cut-point $\leftarrow v$;
4: $o$.entry-graph $\leftarrow$ entry-graph($v$);
5: let $C = \{G_1, G_2, \cdots\}$ be the connected components after removing $v$ and entry-graph($v$) from $G$;
6: **for all** $G_i \in C$ **do**
7:   **if** cut-points $\bigcap V(G_i) \neq \emptyset$ **then**
8:     $o' \leftarrow$ cptree-construct($G_i$, cut-points $\bigcap V(G_i)$);
9:     $o$.subnodes $\leftarrow o$.subnodes $\bigcup \{o'\}$;
10:   **else**
11:     $o$.left-graph $\leftarrow o$.left-graph $\bigcup G_i$;
12: **return** $o$;

**Algorithm 10** cp-search($G, o, k$)

**Input**: Graph $G$, the cpnode $o$ of the cptree, the top-$k$ value.
**Output**: The search result $o$.result$_0$ and $o$.result$_1$.

1: **for all** $o' \in o$.subnodes **do**
2:   cp-search($o'$);
3: **for** $i = 0$ **to** $1$ **do**
4:   **if** $i = 1$ **then**
5:     mark($o$.cut-point.adj($G$));
6:   $o$.result$_i \leftarrow$ div-cut(remove-mark($o$.left-graph), $k$);
7:   **for all** $o' \in o$.subnodes **do**
8:     $D \leftarrow \emptyset$;
9:     **for** $j = 0$ **to** $1$ **do**
10:       **if** $j = 1$ **and** $i = 1$ **and** $o'$.cut-point $\in o$.cut-point.adj($G$) **then**
11:         **break**;
12:       **if** $j = 1$ **then**
13:         mark($o'$.cut-point.adj($G$));
14:       $D' \leftarrow$ div-cut(remove-mark($o'$.entry-graph), $k$);
15:       $D' \leftarrow o'$.result$_j \oplus D'$;
16:       $D \leftarrow D \otimes D'$;
17:       **if** $j = 1$ **then**
18:         unmark($o'$.cut-point.adj($G$));
19:     $o$.result$_i \leftarrow o$.result$_i \oplus D$;
20:   **if** $i = 1$ **then**
21:     update $o$.result$_i$ by adding node $o$.cut-point into every solution of $o$.cut-point;
22:     unmark($o$.cut-point.adj($G$));

**Example 4** *Fig. 8 shows a sample diversity graph. In the graph, $w_1$ can be removed since there exits a node $w_2$ that connects $w_1$, and every node that $w_2$ connects is connected to $w_1$. After compression, the new graph is shown in Fig. 9. The* cptree *of the new graph is shown on the left most part of Fig. 11, where there are 3 nodes, $w_2$, $w_4$ and $w_5$ with root $w_2$. The $w_4$.*entry-graph *is $G_1'$ which is marked on Fig. 9, and the left graph of $w_4$ is a graph that contains only one node $w_3$.*

**Solution Overview:** The cut point based solution is outlined in Algorithm 8. Similar to Algorithm 7, it first decomposes the diversity graph $G$ into disconnected components (line 2). For each component $G_i$, we first compress it by removing nodes based on Lemma 7 (line 4). If there are no cutting points, we simply search $G_i$ using div-astar (Algorithm 4). Otherwise, we construct the cptree with root $o$ for $G_i$ and search on the cptree from root node $o$ to calculate $o$.result$_0$ and $o$.result$_1$ (line 8-9). $o$.result$_0$ and $o$.result$_1$ are combined using the $\otimes$ operator since they are for the same subset of nodes (line 10). The results for different components are combined using the $\oplus$ operator since they are for different subset of nodes in $G$ (line 11). We discuss how to construct the cptree and how to search the cptree below.

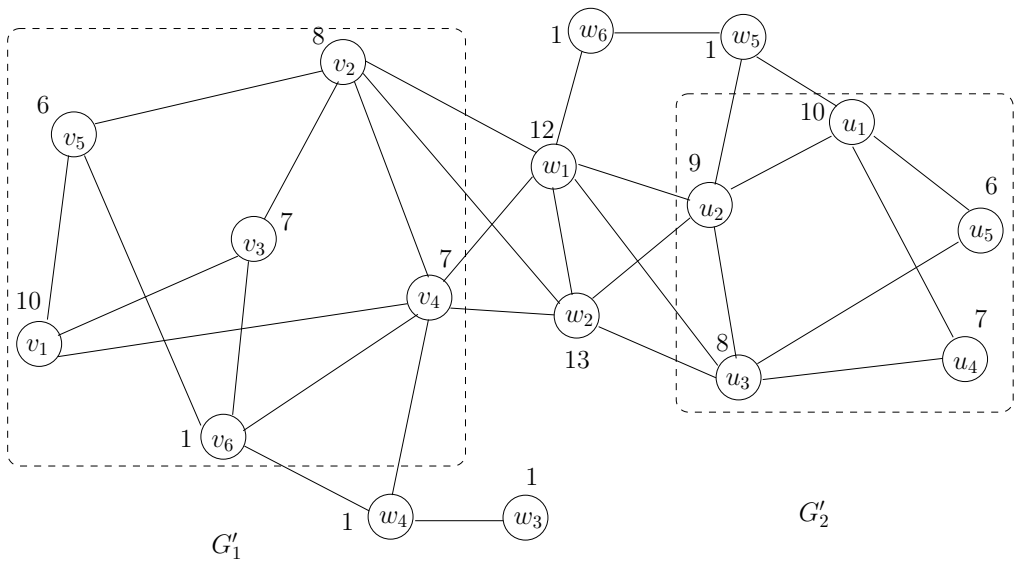

**Figure 8: A sample diversity graph**

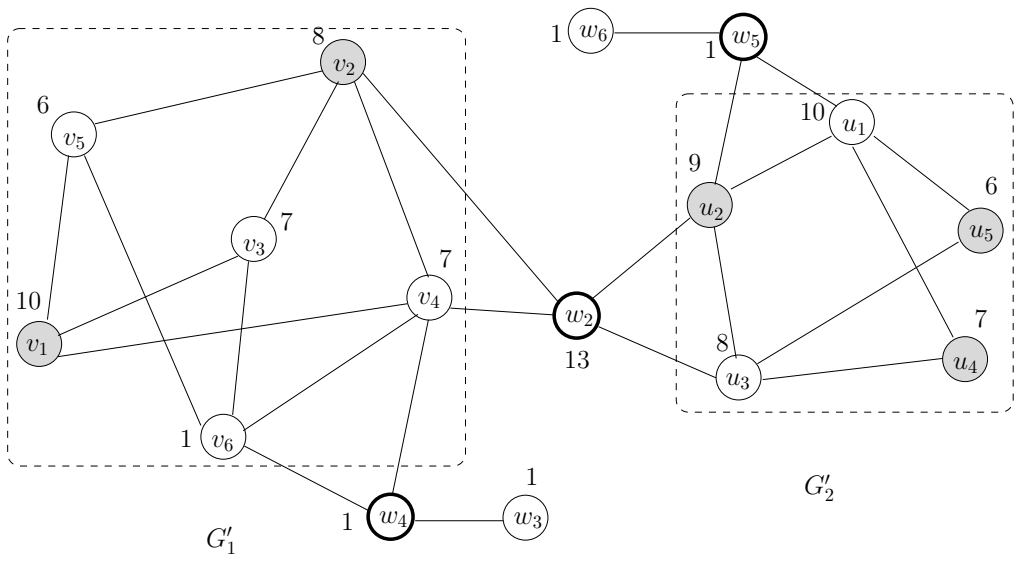

**Figure 9: Compressed diversity graph**

**Constructing the** cptree**:** Given a diversity graph $G$, the cptree for $G$ is constructed as follows. First, the set of cut points, cut-points, is computed using the Tarjan's algorithm with linear time w.r.t. the size of $G$. Then each subtree of cptree is constructed recursively based on a certain subgraph $G'$. The root node $v$ of the subtree is selected as follows: if $v$ is the root of the whole cptree, $v$ is the node in cut-points, such that after removing $v$, the maximum component of $G$ is minimized. Otherwise, $v$ is selected as the node in cut-points, such that after removing $v$ from $G'$, the size of the component that is connected to $v$'s farther node in the original graph $G$, denoted as entry-graph($v$), is maximized. For other components in $G'$ after removing $v$, they can be divided into two categories. The first category includes components with no node in cut-points. Such components are added to the left-graph of $v$. The other category includes components with at least one node in cut-points. Each of such components is considered as a subtree of $v$ in cptree and is created recursively. The algorithm for constructing the cptree is shown in Algorithm 9.

**Searching the** cptree**:** The aim of searching the cptree is to compute $o$.result$_0$ and $o$.result$_1$ for every node $o$ on the cptree in a bottom-up fashion. For a certain node $o$ on the cptree, suppose the result$_0$ and result$_1$ for all $o$'s subnodes have been computed, we need to compute $o$.result$_0$ and $o$.result$_1$. The algorithm to search the cptree is shown in Algorithm 10.

We explain Algorithm 10 using an example. Fig. 10 shows a cptree with 3 nodes, $o_{12}$, $o_{34}$ and $o_{24}$ connecting 4 graphs $G_1$, $G_2$, $G_3$ and $G_4$. $G_{34}$ consists of $G_3$, $G_4$ and $o_{3,4}$. $G_{12}$ consists of $G_1$, $G_2$ and $o_{12}$, and $G$ consists of $G_{12}$, $G_{34}$ and $o_{24}$. For simplicity, in this example, we use the graph itself to denote the search result on the graph. For a cutting point $o$ on a graph $G$, we use $G$.include$_o$ to denote the optimal solution on $G$ that $o$ is included, and use $G$.exclude$_o$ to denote the optimal solution on $G$ that $o$ is excluded. Suppose $G_3$.include$_{o_{34}}$, $G_3$.exclude$_{o_{34}}$, $G_1$.include$_{o_{12}}$, and $G_1$.exclude$_{o_{12}}$ have been computed. We show how to compute $G$.include$_{o_{24}}$ and $G$.exclude$_{o_{24}}$.

**Computing** $G$.exclude$_{o_{24}}$**:** It is the case for $i = 0$ in Algorithm 10 (line 3). Since $o_{24}$ is excluded, we have $G$.exclude$_{o_{24}} = G_{34} \oplus$

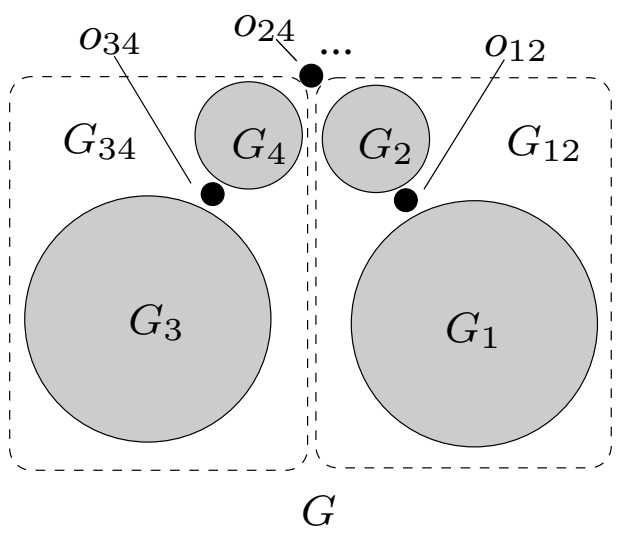

**Figure 10:** cptree **search example**

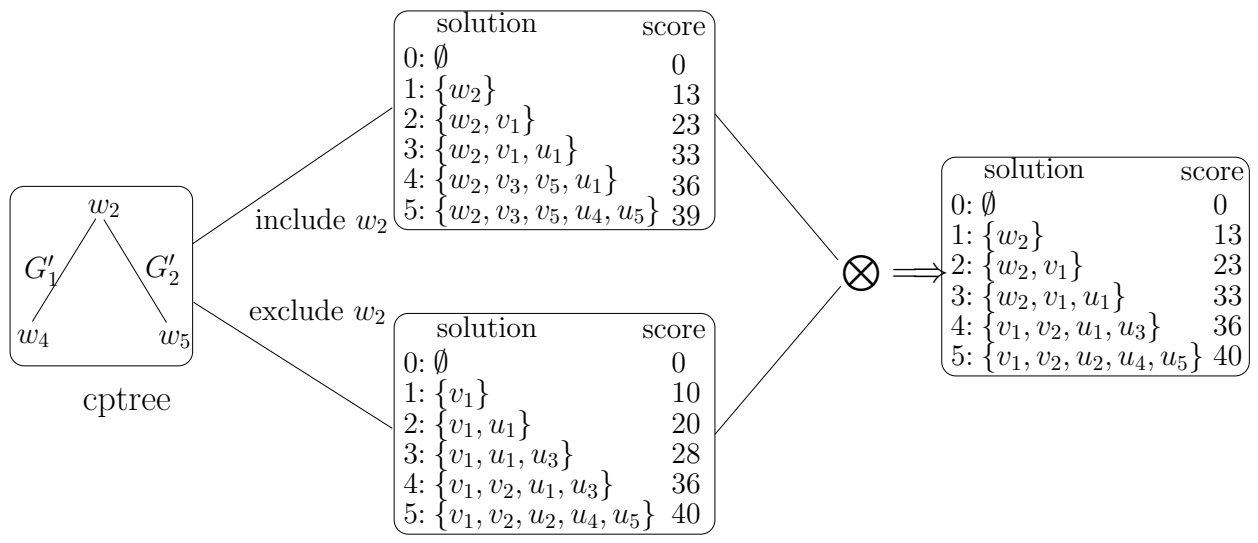

**Figure 11: Search the** cptree

$G_{12}$ (line 19 in the for loop from line 7). We now discuss how to compute $G_{12}$, and $G_{34}$ can be computed in a similar way. There are two situations:

1. $o_{12}$ is excluded. It is the case for $j = 0$ (line 9). In such a situation, we can compute $G'_{12} = G_1.\mathsf{exclude}_{o_{12}} \oplus G_2$ (line 15).

2. $o_{12}$ is included. It is the case for $j = 1$ (line 9). In such a situation, we can compute $G''_{12} = G_1.\mathsf{include}_{o_{12}} \oplus (G_2 - o_{12}.\mathsf{adj}(G))$ (line 15), where $G_2 - o_{12}.\mathsf{adj}(G)$ is to remove the adjacent nodes of $o_{12}$ from $G_2$ (line 13).

After computing 1 and 2, $G_{12}$ can be computed as $G_{12} = G'_{12} \otimes G''_{12}$ (line 16).

**Computing** $G.\mathsf{include}_{o_{24}}$**:** It is the case for $i = 1$ in Algorithm 10 (line 3). Since $o_{24}$ is included, we should remove nodes that are adjacent to $o_{24}$ in $G_4$ and $G_2$ (line 5). Thus $G.\mathsf{include}_{o_{24}} = (G_{34} - o_{24}.\mathsf{adj}(G)) \oplus (G_{12} - o_{24}.\mathsf{adj}(G)$. Note that since $o_{24}$ is included, we should also add $o_{24}$ into $G.\mathsf{include}_{o_{24}}$ after computing the $\oplus$ operation (line 21). We now discuss how to compute $(G_{12} - o_{24}.\mathsf{adj}(G))$, and $(G_{34} - o_{24}.\mathsf{adj}(G))$ can be computed in a similar way. There are two situations:

1. $o_{12}$ is excluded. It is the case for $j = 0$ (line 9). In such a situation, we can compute $G'_{12} = G_1.\mathsf{exclude}_{o_{12}} \oplus (G_2 - o_{24}.\mathsf{adj}(G))$ (line 15).

2. $o_{12}$ is included. It is the case for $j = 1$ (line 9). In such a situation, we can compute $G''_{12} = G_1.\mathsf{include}_{o_{12}} \oplus (G_2 - o_{24}.\mathsf{adj}(G) - o_{12}.\mathsf{adj}(G))$ (line 15), where $G_2 - o_{24}.\mathsf{adj}(G) - o_{12}.\mathsf{adj}(G)$ is to remove the adjacent nodes of $o_{12}$ from $G_2 - o_{24}.\mathsf{adj}(G)$ (line 13).

After computing (1) and (2), $(G_{12} - o_{24}.\mathsf{adj}(G))$ can be computed as $(G_{12} - o_{24}.\mathsf{adj}(G)) = G'_{12} \otimes G''_{12}$ (line 16).

From the above discussion, on the cptree, for each node $o'$, $o'$.entry-graph needs to be searched for at most four times depending on whether $o$.cut-point and $o'$.cut-point are included or not, where $o$ is the father node of $o'$ on cptree. For the above example, $G_2$ is searched four times, using $G_2$, $(G_2 - o_{12}.\mathsf{adj}(G))$, $(G_2 - o_{24}.\mathsf{adj}(G))$ and $(G_2 - o_{24}.\mathsf{adj}(G) - o_{12}.\mathsf{adj}(G))$ respectively. There are two more cases that the above example is not considered. First, when $o$.left-graph is not $\emptyset$, the search result on $o$.left-graph should also be combined into $o.\mathsf{result}_0$ and $o.\mathsf{result}_1$ using operator $\oplus$ (line 6). Second, when $o$ and $o'$ are adjacent, and both $o$ and $o'$ are included, it is not a feasible solution (line 10-11).

**Example 5** *For the diversity search graph shown in Fig. 8, suppose $k = 5$, after graph compression (Fig. 9), the* cptree *is shown in the left most part of Fig. 11. For the root node $w_2$, suppose the solutions for its* subnodes *$w_4$ and $w_5$ are computed. The optimal solution $w_2.\mathsf{result}_0$ (exclude $w_2$) is computed by combing two results using operator $\oplus$: (1) the optimal solution for $G'_1 \bigcup$ the subgraph represented by $w_4$, and (2) the optimal solution for $G'_2 \bigcup$ the subgraph represented by $w_5$. When computing (1), we should combine the following two results using the operator $\otimes$: (a) the result by including $w_4$ and (b) the result by excluding $w_4$. (a) can be computed by first removing all nodes from $G'_1$ that are adjacent to $w_4$, and combining the optimal solutions for $G'_1$ and $w_4.\mathsf{result}_1$ using operator $\oplus$. The optimal solution for $G'_1$ can be computed using* div-cut *again. (b) can be computed in a similar way. The optimal solution $w_2.\mathsf{result}_0$ is shown in the middle lower part of Fig. 11, and the optimal solution $w_2.\mathsf{result}_1$ is shown in the middle upper part of Fig. 11. After combining $w_2.\mathsf{result}_0$ and $w_2.\mathsf{result}_1$ using operator $\otimes$, the final solution is shown in the right part of Fig. 11.*

| kfreq | enwiki | reuters |
|---|---|---|
| 1 | historic lake | spokesman |
| 2 | divided areas | net |
| 3 | rating system | april |
| 4 | student high school | billion |
| 5 | community web | march |

**Figure 12: Keyword queries**

# 8. PERFORMANCE STUDIES

We conducted extensive performance studies to test the algorithms proposed in this paper. We implemented three algorithms, denoted div-astar (Algorithm 4), div-dp (Algorithm 7), and div-cut (Algorithm 8) that follow the framework shown in Algorithm 3 with different implementations on div-search-current. All algorithms were implemented in Visual C++ 2008 and all tests were conducted on a 2.8GHz CPU and 2GB memory PC running Windows XP.

We use two real datasets, enwiki[2] and reuters[3]. enwiki includes 11,930,681 articles from the English Wikipedia, and reuters include 21,578 news from Reuters. Given a user keyword query $q$, we search the top-$k$ documents using the TF*IDF score normalized by the length of the corresponding document, which is defined as follows for each document $d$.

$$\mathsf{score}(q, d) = \frac{\sum_{q_i \in q} tf(q_i, d) \times idf(q_i)}{\sqrt{len(d)}} \quad (3)$$

where $tf(q_i, d)$ is term frequency of keyword $q_i$ in $d$, $idf(q_i)$ is the inverted document frequency for keyword $q_i$, which is defined as $idf(q_i) = \log \frac{|D|}{|\{d \in D : q_i \in d\}|+1}$ for the dataset $D$, and $len(d)$ is the total number of words in $d$. Given any two documents $d_1$ and $d_2$, suppose all stop words have been removed from $d_1$ and $d_2$, the similarity for $d_1$ and $d_2$ is defined as follows, based on the weighted Jaccard distance.

[2]http://en.wikipedia.org/wiki/Wikipedia:Database_download

[3]http://kdd.ics.uci.edu/databases/reuters21578/

$$\mathsf{sim}(d_1, d_2) = \frac{\sum_{w \in d_1 \bigcap d_2} idf(w)}{\sum_{w \in d_1 \bigcup d_2} idf(w)} \quad (4)$$

where $d_1 \bigcap d_2$ is the multi-set of words that appear in both $d_1$ and $d_2$, and $d_1 \bigcup d_2$ is the multi-set of words that appear in either $d_1$ or $d_2$.

For enwiki, we test the scalability for keyword queries with multiple keywords, where the results for each keyword are sorted according to the scores and stored in the inverted index. The results for all keywords are aggregated using the threshold algorithm [8]. For reuters, we test the scalability for keyword queries where each keyword query only contains one keyword. The results are output incrementally by sequentially scanning the inverted index for the keyword. For each testing, we record the processing time and the peak memory consumption. The processing time/peak memory consumption is the total time/peak memory consumed in searching the diversified top-$k$ results. When all the 2GB memory is used up, the algorithm cannot compute the diversified top-$k$ results. We use INF to denote such a situation.

For each dataset, we vary 3 parameters, $k$, $\tau$ and kfreq. $k$ is the top-$k$ value, $\tau$ is the similarity threshold, and kfreq is the average keyword frequency for the corresponding query. For each dataset, we select representative queries with different keyword frequencies as follows. After removing all the stop words, we set the maximum keyword frequency among all keywords as $\pi$, and divide the keyword frequency range between 0 and $\pi$ into 5 partitions, namely, $\pi/5$, $2\pi/5$, $3\pi/5$, $4\pi/5$ and $\pi$. For simplicity, we say a keyword has frequency $p(p \in \{1, 2, 3, 4, 5\})$, iff its frequency is between $(p-1) \cdot \pi/5$ and $p \cdot \pi/5$. We also set two groups of $k$ values. The small $k$ values and the large $k$ values. Since div-astar is not suitable to be processed when $k$ is large, in the large $k$ value group, we only compare the two algorithms div-dp and div-cut. For enwiki, $k$ is selected from $\{40, 80, 120, 160, 200\}$ with default value 120 for small $k$ values and selected from $\{500, 700, 900, 1300, 2000\}$ with default value 900 for large $k$ values. $\tau$ is selected from $\{0.4, 0.5, 0.6, 0.7, 0.8\}$ with default value 0.6, and kfreq ranges from 1 to 5 with default value 3. For reuters, $k$ is selected from $\{60, 80, 100, 110, 120\}$ with default value 100 for small $k$ values and selected from $\{500,$ $700,$ $900,$ $1300,$ $2000\}$ with default value 900 for large $k$ values. The small $k$ values selected in reuters are different from those in enwiki because div-astar can hardly handle queries when $k$ is as large as 200 in reuters. $\tau$ is selected from $\{0.4, 0.5, 0.6, 0.7, 0.8\}$ with default value 0.6, and kfreq ranges from 1 to 5 with default value 3. When varying a certain parameter, the values for all the other parameters are set to their default values. The set of keywords with different kfreq are shown in Fig. 12.

**Exp-1 (Test enwiki):** The testing results on the enwiki dataset when varying $k$ are shown in Fig. 13. Fig. 13 (a) and Fig. 13 (b) show the processing time and memory consumption when $k$ is small. When $k$ increases, the processing time for all the three algorithms div-cut, div-dp, and div-astar increase. div-astar increases sharply and div-dp and div-astar keep stable. When $k$ reaches 200, div-astar takes more than 200 seconds and consumes more than 200MB memory while both div-cut and div-astar take less than 0.1 seconds and consumes less than 10KB memory. The processing time and memory consumption for large $k$ values are shown in Fig. 13 (c) and Fig. 13 (d) respectively. When $k$ increases, the time and memory consumption for div-dp increase sharply while the time and memory consumption for div-cut increase slowly. This

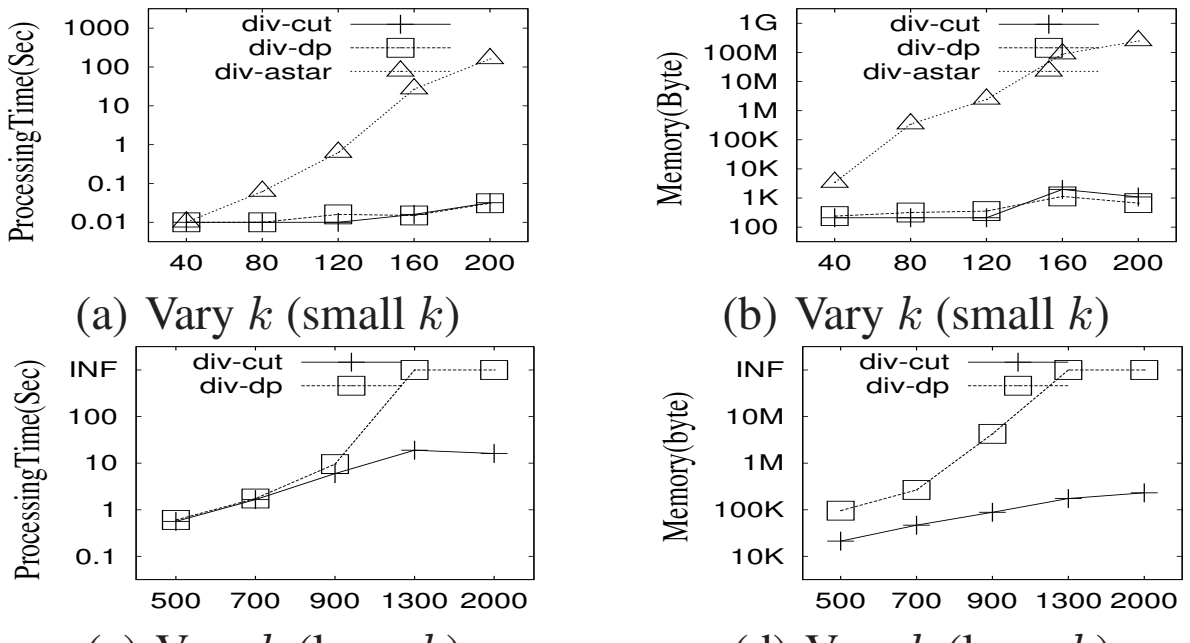

(a) Vary $k$ (small $k$) (b) Vary $k$ (small $k$)
(c) Vary $k$ (large $k$) (d) Vary $k$ (large $k$)

**Figure 13: Vary $k$ (enwiki)**

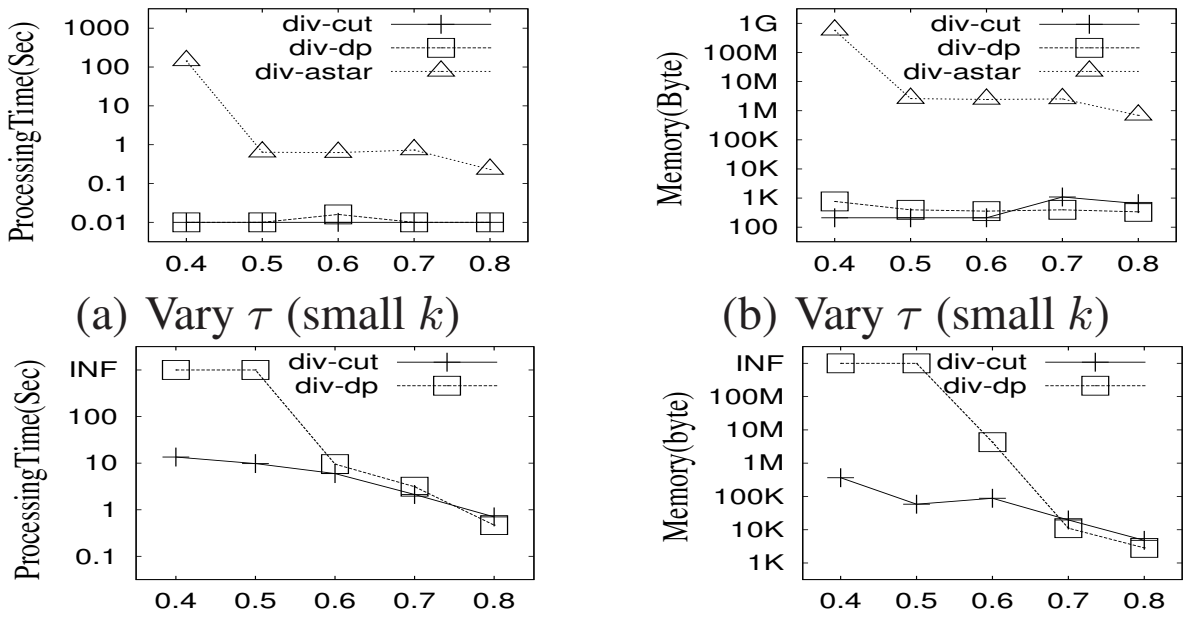

(a) Vary $\tau$ (small $k$) (b) Vary $\tau$ (small $k$)
(c) Vary $\tau$ (large $k$) (d) Vary $\tau$ (large $k$)

**Figure 14: Vary $\tau$ (enwiki)**

is because, when $k$ increases, the size of the largest component for the diversity graph increases, but it can still be decomposed into relatively smaller subgraphs using cut points. When $k$ reaches $2,000$, div-dp cannot generate the result and div-cut can compute the optimal solution within 15 seconds using less than 200KB memory.

When $\tau$ varies from 0.4 to 0.8, the testing results on the enwiki dataset are shown in Fig. 14. As shown in Fig. 14 (a) and Fig. 14 (b), for small $k$, when $\tau$ increases, the processing time and memory consumption for div-astar decrease, and the time and memory consumption for div-cut and div-dp keep stable. This is because when $\tau$ is large, a large number of search results are not similar to each other. Thus in the div-astar search, the estimated upper bound is tight and thus the algorithm stops early. For large $k$ values, as shown in Fig. 14 (c) and Fig. 14 (d), when $\tau$ increases, the time and memory consumption for both div-cut and div-dp decrease. When $\tau$ is 0.4, div-dp cannot generate a result because when $\tau$ is small, a lot of search results are similar to each other, and thus the largest component for the diversity graph is large. div-cut can compute the optimal solution within 15 seconds using less than 1MB memory.

Fig. 15 shows the testing results on the enwiki dataset when varying kfreq. When kfreq increases, both the processing time and memory consumption do not have an obvious trend to increase or decrease. This is because whether two search results are similar to each other is not dependent largely on the keyword frequency for the query. Fig. 15 (a) and Fig. 15 (b) show the processing time and memory consumption for small $k$ values. div-cut and div-dp have similar performance and div-astar is more than 100 times slower and consumes 1000 times more memory, comparing to div-cut and div-dp in all cases. The processing time and memory consumption for large $k$ values when varying kfreq are shown in Fig. 15 (c) and Fig. 15 (d) respectively. div-dp is more than 2 times slower and consumes 10 times more memory comparing to div-cut in all cases. When kfreq $=2$, div-dp cannot generate a result and div-cut can

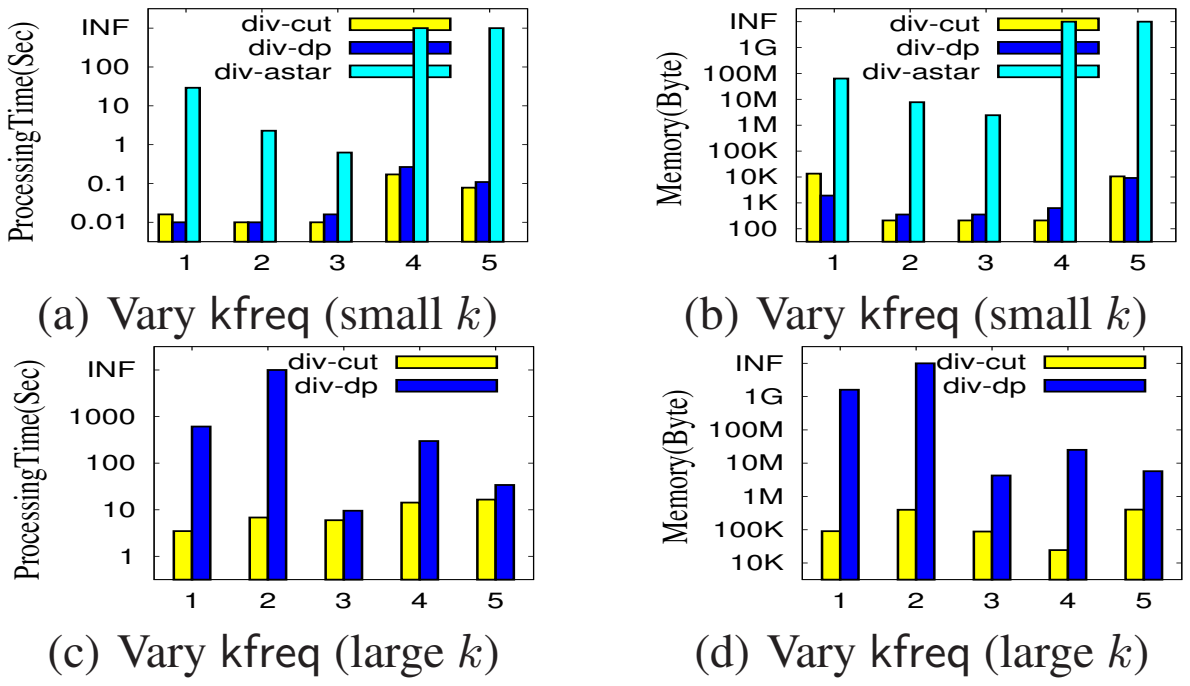

(a) Vary kfreq (small $k$) (b) Vary kfreq (small $k$)

(c) Vary kfreq (large $k$) (d) Vary kfreq (large $k$)

**Figure 15: Vary** kfreq (enwiki)

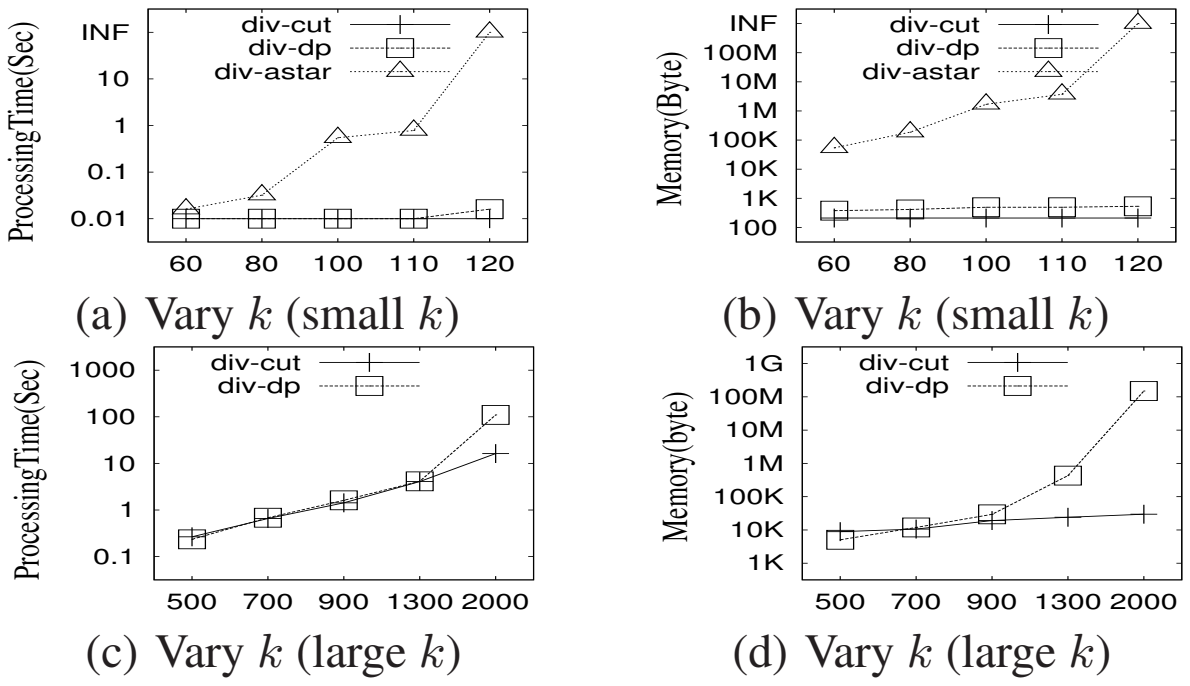

(a) Vary $k$ (small $k$) (b) Vary $k$ (small $k$)

(c) Vary $k$ (large $k$) (d) Vary $k$ (large $k$)

**Figure 16: Vary** $k$ (reuters)

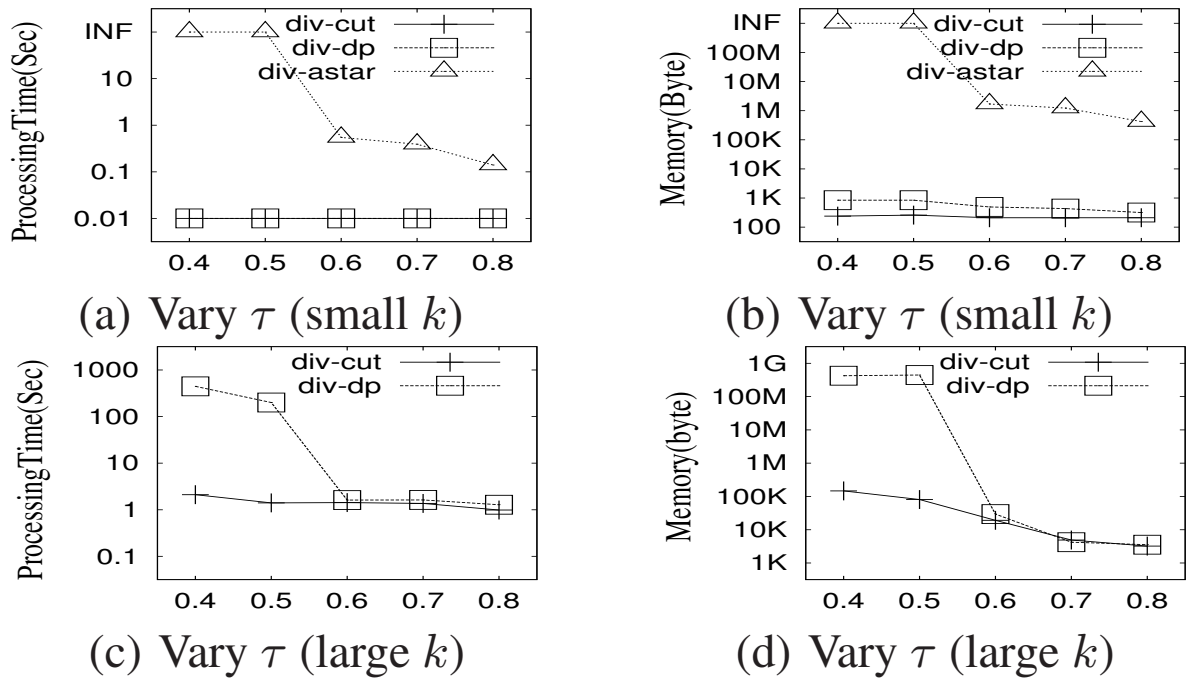

(a) Vary $\tau$ (small $k$) (b) Vary $\tau$ (small $k$)

(c) Vary $\tau$ (large $k$) (d) Vary $\tau$ (large $k$)

**Figure 17: Vary** $\tau$ (reuters)

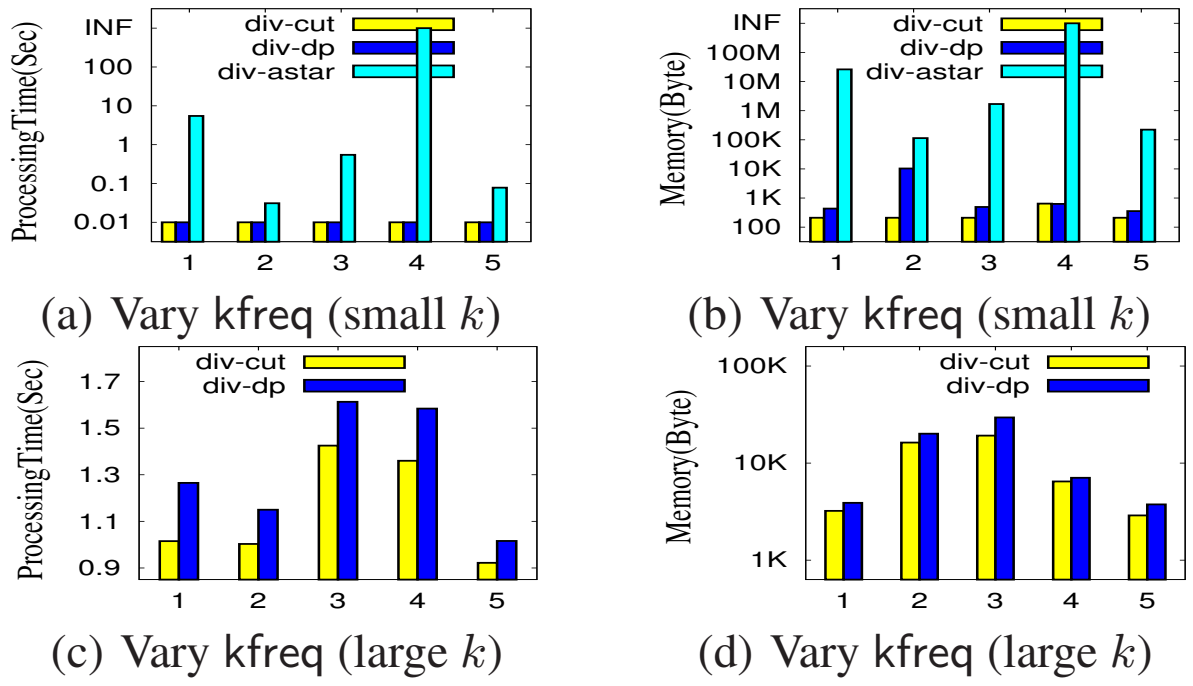

(a) Vary kfreq (small $k$) (b) Vary kfreq (small $k$)

(c) Vary kfreq (large $k$) (d) Vary kfreq (large $k$)

**Figure 18: Vary** kfreq (reuters)

finish in less than 10 seconds using less than 500KB memory.

**Exp-2 (Test** reuters**):** The testing results when varying $k$ in the reuters dataset are shown in Fig. 16. Fig. 16 (a) and Fig. 16 (b) show that, for small $k$ values, when increasing $k$, the processing time and memory consumption for div-cut and div-dp keep stable and the time/memory consumption for div-astar increase sharply. When $k$ reaches 120, div-astar cannot generate a result and div-cut and div-dp can finish in less than 0.1 seconds using less than 1KB memory. For large $k$ values, as shown in Fig. 16 (c) and Fig. 16 (d), when $k$ is less than 900, div-cut and div-dp have almost the same performance. When $k = 500$, div-dp even consumes smaller memory than div-cut. It is because when $k$ is small, the components for the diversity graph are all small, thus both div-cut and div-dp consumes small memory, but div-cut needs extra space to put the cptree. When $k$ is as large as 2000, div-dp is 10 times slower than div-dp and consumes more than 1000 times more memory.

The curves for reuters when varying $\tau$ are shown in Fig. 17. As shown in Fig. 17(a) and Fig. 17(b), for small $k$ values, when $\tau$ increases, the time/memory consumption for div-astar decrease sharply and div-cut and div-dp keep stable. When $\tau \leq 0.5$, the div-astar algorithm cannot generate a result while div-astar and div-cut can compute the optimal solution within 0.01 seconds using more than 1KB memory. The results for large $k$ values are shown in Fig. 17 (c) and Fig. 17 (d). When $\tau$ is larger than 0.6, div-cut and div-dp have almost the same performance. When $\tau$ decreases, the time/memory consumption for div-dp increase sharply and div-cut keeps stable. For $\tau = 0.4$, div-dp can compute the optimal solution in more than 500 seconds using more than 500MB memory, while div-cut can compute the optimal solution in less than 5 seconds using less than 200KB memory.

Fig. 18 shows the testing results when varying kfreq in the reuters dataset. Again, the time/memory consumption does not have an obvious trend to increase or decrease. For small $k$ values, as shown in Fig. 18 and Fig. 18, div-cut and div-dp have similar processing time, but div-dp consumes more memory than div-cut. div-astar is much slower and consumes much more memory than div-cut and div-dp in all cases. For large $k$ values, as shown in Fig. 18, the gap between div-dp and div-cut is not as large as those in the enwiki dataset. It is because in the enwiki dataset, the number of documents is large, and thus documents that fall into the same category can be similar to each other with high probability, and in the reuters dataset, the number of documents is small, and thus the probability that two documents are similar to each other is small. div-cut is faster and consumes smaller memory than div-dp in all cases.

# 9. RELATED WORK

In general, our problem of finding diversified top-k results is related to three problems in the literature, namely, traditional top-k query, diversified top-k query, and maximum weight independent set.

**Traditional top-k query**: In a top-k query, each result is associated with a score and the $k$ results with largest scores are reported as the top-k results. General techniques to answer top-k queries follow into two categories.

In the first category, all the results define a solution space. The approach recursively partitions the solution space into subspaces based on the best result in the current subspace, and the next best result is the one with largest score among the best result in each subspace. Therefore, the top-k results are generated one-by-one using Lawler's procedure [16], and the approaches based on it are in [15, 20, 14]. Lawler [16] proposes a general procedure for computing top-k results to discrete optimization problems and also discusses its application to k shortest path problem. Based on Lawler's procedure, Kimelfeld and Sagiv [15] study how to find top-k steiner trees, Qin et al. [20] focus on finding top-k communities, Kargar and An [14] find r-cliques in a graph.

In the other category, the results are generated in a heuristic order, and an upper bound score is computed for all the ungenerated results. The algorithm stops if the scores of current top-k results are no smaller than the upper bound score. A detailed survey can be found in [13]. The general setting is that, one ranked list is defined for each query feature, and the score of each result is an aggreation of scores corresponding to query features, which is a classical setting in Information Retrieval. The most inflential algorithm is proposed by Fagin et al. [7, 9], which considers both random access and/or sequential access of the ranked lists. Other works consider the scenario that only sequential accesses of the ranked lists are allowed [12, 17].

**Diversified top-k query**: In the traditional top-k queries, it returns results based on their relevance scores only. More and more works propose to take the diversity (or redundancy) into consideration to return a more satisfied result list for user queries [2, 1, 23, 22, 11]. The problems studied about diversity expand a wide variety of spectrum, e.g., diversified keyword search in documents [2], diversified prestige node finding in information networks [18], diversified keyword query recommendation [23], diversified document monitoring in information filtering system [22], diversified keyword query interpretation over structured databases [6], and diversified keyword search over graphs [11], and so on.

In general, for a diversity-aware query, the results are returned in an ordered list and a redundancy value is computed for each result based on the content of results preceding it. Then, a usefulness score is computed by combining the relevance score and redundancy value, and results are ranked with respect to the usefulness score [2, 1, 6, 11]. The approaches to find top-k diversified results by usefulness scores generally consist of two steps, which first computes a top-$l$ ($l > k$) results based on the relevance score only and then reranks the $l$ results based on the usefulness score using a greedy algorithm [6, 1, 11, 5]. Considering the efficiency aspect, Angel and Koudas [2] propose a one-step approach to answering diverisfied top-k queries. They couple their algorithm with the threshold algorithm which is designed for traditional top-k query [9]. An upper bound usefulness score is computed for the non-retrieved documents, the current $k$ documents with largest scores are the top-k results if their scores are no smaller than the upper bound computed.

Different from the above works, Zhang et al. [22] treat the redundancy of a document with respect to a set of relevant documents as a binary value, i.e., a document is either redundant or should be reported as relevant. Mei et al. [18] find the top-k diversified prestige nodes in information networks using vertex-reinforced random walk. Zhu et al. [23] recommend top-k diversified relevant queries using a manifold based approach.

**Maximum weight independent set**: Our problem can be viewed as an instance of finding maximum weight independent set constrained with size $k$, which is NP-hard [10]. The problems of finding maximum weight independent set, maximum weight clique, and minimum weight vertex cover are all correlated, and these problems are hard to approximate. Therefore, very few attempts have been done in the literature to find exact solutions, except the branch-and-bound methods [21, 3, 19, 4]. However, these works do not consider the size constraint $k$ as introduced in our problem. Also, in our problem, the diversity graph is not toally materalized.

# 10. CONCLUSION

In this paper, we study the diversified top-$k$ search problem, that take both the scores of results and diversity into consideration. We formally define the problem using the similarity of search results themselves. We propose a framework, such that most existing solutions that handle top-$k$ query processing with early stop can be used in our framework to handle diversified top-$k$ search by applying three new functions, namely, a sufficient stop condition sufficient(), a necessary top condition necessary() and an diversified top-$k$ search algorithm div-search-current() to search on the current result set. We study all the three functions in details and give three algorithms for div-search-current(). We conducted extensive performance studies to show the performance of our algorithms.

**Acknowledgment**: The work was supported by grant of the Research Grants Council of the Hong Kong SAR, China No. CUHK 419109.